\documentclass[pre,aps,10pt,superscriptaddress,tightenlines,showpacs,nofootinbib,notitlepage,longbibliography]{revtex4-2}
\usepackage[T1]{fontenc}
\usepackage[utf8]{inputenc}
\usepackage{lmodern}
\usepackage{cancel}
\usepackage{graphicx,amssymb,amsmath,epsf,bm}
\usepackage[normalem]{ulem}
\usepackage{stackrel}
\usepackage{color}
\usepackage{upgreek}

\usepackage{latexsym,amsmath,amsfonts,tabularx,amssymb,bm,empheq}
\usepackage{graphicx,epstopdf}
\usepackage{xspace}

\usepackage{hyperref}
\hypersetup{
  citecolor = {green},
  colorlinks = {true}, 
  urlcolor = {blue} 
}
 \usepackage[capitalise,]{cleveref}

\newcommand{\nnl}{\nonumber \\}
 
\def\q{\mathbf{q}}
\def\p{\mathbf{p}}

\def\n{\mathbf{n}}
\def\x{\textbf{x}}
\def\y{\textbf{y}}
\def\z{\textbf{z}}

\def\pc{\boldsymbol{\mathsf p}}
\def\qc{\boldsymbol{\mathsf q}}

\newcommand{\zero}{\mathbf{0}}

\newcommand{\hphi}{\hat{\phi}}

\newcommand{\cGamma}{\check\Gamma}
\newcommand{\cU}{I}
\newcommand{\cZ}{\check Z}
\newcommand{\cG}{\check G}
\newcommand{\cSigma}{\check \Sigma}
\newcommand{\cDelta}{\check \Delta}

\newcommand{\DD}{\mathcal{D}}

\def\simge{\mathrel{%
   \rlap{\raise 0.511ex \hbox{$>$}}{\lower 0.511ex \hbox{$\sim$}}}}
\def\simle{\mathrel{
   \rlap{\raise 0.511ex \hbox{$<$}}{\lower 0.511ex \hbox{$\sim$}}}}

\def\simle{\mathrel{
   \rlap{\raise 0.511ex \hbox{$<$}}{\lower 0.511ex \hbox{$\sim$}}}}

\def\simge{\mathrel{%
    \rlap{\raise 0.511ex \hbox{$>$}}{\lower 0.511ex \hbox{$\sim$}}}}

\usepackage{subfigure}
\usepackage{color}
\usepackage{tabularx}

\definecolor{CornFlowerBlue}{rgb}{0.39, 0.58, 0.93}
\definecolor{ForestGreen}{rgb}{0.13, 0.55, 0.13}
\definecolor{WildStrawberry}{rgb}{1.0, 0.26, 0.64}

\begin{document}

\title{Constraint effective action and critical correlation functions at fixed magnetization}

\author{F\'elix Rose}
\affiliation{Laboratoire de Physique Th\'eorique et Mod\'elisation,\\
CY Cergy Paris Universit\'e, CNRS, F-95302 Cergy-Pontoise, France}

\affiliation{Institut f\"ur Theoretische Physik, Universit\"at Heidelberg, 69120 Heidelberg, Germany}
\author{Adam Ran\c con}
\affiliation{Univ. Lille, CNRS, UMR 8523 -- PhLAM -- Laboratoire de Physique
des Lasers Atomes et Mol\'ecules, F-59000 Lille, France}
\affiliation{Institut Universitaire de France}

\author{Ivan Balog}
 \affiliation{Institute of Physics, Bijeni\v{c}ka cesta 46, HR-10001 Zagreb, Croatia}

\date{May 6, 2026}

\begin{abstract}
We present an extension of the functional renormalization group (FRG) framework developed to compute critical probability distributions of the order parameter to momentum‐dependent observables. Focusing on the constraint effective action at fixed magnetization for the Ising universality class, we derive its exact flow equations and solve them at the second order of the derivative expansion (DE2). We solve these flow equations numerically for two‐ and three‐dimensional systems, extract universal rate functions and momentum‐dependent correlation functions, and benchmark them against Monte Carlo simulations. In three dimensions, we recover the rate function and accurately reproduce the first few Fourier modes of the constrained correlation function and demonstrate the convergence of the method. In two dimensions, the lowest-order approximations such as local potential approximation (LPA) fail, and it is required to consider at least the DE2 to describe the critical point. Our results are in qualitative agreement with the numerics. We confirm the robustness of the FRG approach for calculating both zero‐ and finite‐momentum critical observables at fixed magnetization.
\end{abstract}

\maketitle


\section{Introduction}

Continuous phase transitions exhibit universal behavior, characterized by scaling laws of correlation functions and critical exponents, independent of microscopic details. A striking manifestation of this universality arises in the probability distribution function (PDF) of macroscopic observables, such as the order parameter. Near criticality, when the correlation length diverges, the central limit theorem, valid when variables are weakly correlated, fails due to strong, system-wide correlations. Instead, the PDF of the order parameter adopts a universal, non-Gaussian form described by a scaling function called the rate function \cite{Bouchaud1990}.
The study of such PDFs has been the subject of intense theoretical and numerical, as well as experimental, works.
On the theoretical side, scaling theory and field-theoretical analyses within perturbative renormalization group (RG) have clarified how critical fluctuations lead to universal, non-Gaussian PDFs in the Ising and $O(N)$ universality class \cite{Bruce1979,Rudnick1985,Eisenriegler1987,Brankov1989,Bruce1992,Hilfer1995,Esser1995,Chen1996,Bruce1997,Rudnick1998,Portelli_02,sahu_generalization_2025,sahuCriticalProbabilityDistributions2025a,Sahu_2026}. 
These theoretical studies were confirmed by extensive numerical Monte Carlo (MC) simulations, which quantitatively confirmed finite-size scaling, universality of the full distribution, and the structure of its tails \cite{Binder1981,Binder1981a,kuti_supercomputing_1988,Nicolaides1988,Tsypin1994,Tsypin2000,Hilfer2003,Hilfer2005,Malakis2014,Xu2020}. 
More recently, a functional renormalization group (FRG) approach has been used to revisit the problem, providing a systematic framework to compute critical PDFs \cite{Balog2022,rancon_probability_2025,balog_universal_2025}.
Experimentally, universal non-Gaussian fluctuations of macroscopic observables have been reported in a variety of settings, ranging from turbulence and interfacial growth to equilibrium critical points in ultracold atoms
\cite{BHP_1,joubaud_experimental_2008,Takeuchi2010,allemand_observation_2025}.

For the paradigmatic case of the Ising model at its critical temperature $T_c$, the PDF of the magnetization density $\hat s = L^{-d}\sum_i \hat s_i$ (with $\hat s_i$ local spins, $L$ system size, and $d$ spatial dimension) scales exponentially with system volume, $P_L(\hat s = s) \sim e^{-L^d I(s)}$. The scaling hypothesis predicts that the rate function $I(s)$ is universal, governed by the anomalous dimension $\eta$, and a function of the rescaled variable $\tilde s = L^{(d-2+\eta)/2} s$. Moreover, away from criticality, universality persists more generally, defining an entire family of universal scaling functions parameterized by the ratio $L/\xi_{\infty}$, where $\xi_\infty$ is the bulk correlation length.

The connection between the rate function, scaling, and universality naturally points towards the renormalization group, the fundamental theoretical framework describing critical phenomena. Historically, this link has been acknowledged since the early stages of RG theory \cite{JonaLasino1975,Gallavotti1974,Gallavotti1975}, yet it was only recently explicitly elucidated within the context of the FRG \cite{Balog2022}. The FRG formalism, built upon Wilsonian RG applied at the level of the effective action (the Legendre transform of the free energy), provides a systematic and nonperturbative method to address critical phenomena \cite{Dupuis2021}.

In this context, the rate function can be computed via the \emph{constraint effective action} \cite{Fukuda1975,ORaifeartaigh1986,hasenfratz_study_1989,Gockeler1991,gockeler_constraint_1991,Mukaida1996,Balog2022}, an analog of the standard effective action specifically designed to extract the rate function. This construction highlights the conceptual and practical distinctions between the rate function and the RG fixed point potential, emphasizing that, unlike the fixed point potential, the rate function is a universal function that explicitly depends on the system size and remains independent of the RG scheme. Within this framework, Refs.~\cite{Balog2022,rancon_probability_2025} computed the rate function for the three-dimensional  Ising and O($N$) model using the simplest FRG approximation, the local potential approximation (LPA), obtaining good agreement with MC simulations.

However, several aspects remain unresolved, which we address in this paper. First, the LPA is a relatively crude approximation that neglects field renormalization, imposing a vanishing anomalous dimension $\eta$. We thus aim to test the convergence of the FRG by pushing the calculation to the second-order derivative expansion (DE2). This approach enables us to determine error bars for the rate function. It also makes it possible to compute correlation functions subject to the global constraint of a fixed order parameter, observables that are accessible in MC simulations.

Finally, we explore the applicability of our formalism to the two-dimensional Ising model. In this case, many properties are known rigorously.
However, the moment generating functional of the order-parameter PDF is directly related to the partition function in a homogeneous magnetic field, presenting a theoretical challenge that remains open. Currently, rigorous results are limited to the tail of the PDF \cite{Camia16}. In two dimensions, the large anomalous dimension  ($\eta = 1/4$) renders the LPA inadequate and the DE2 is needed.
We compare our theoretical predictions with MC simulations in two and three dimensions, achieving good agreement and underscoring the robustness and precision of the FRG.

The rest of this paper is organized as follows. In \cref{sec:const}, we introduce the constraint effective action and derive some of its properties. We extend the FRG formalism to the constraint effective action in \cref{sec:FRG} and describe approximation schemes based on the derivative expansion to second order. In \cref{sec:critical}, the critical constraint effective action of the two- and three-dimensional Ising model is computed, and we assess the convergence of the method. Comparisons to MC simulations are presented in \cref{sec:MC}, while our conclusions are given in \cref{sec:concl}.

\section{Constraint effective action \label{sec:const}}

We seek to compute observables of a theory while imposing a constraint on its order parameter. The formalism to do so relies on the constraint effective action $\cGamma$. Akin to the effective action in the absence of a constraint, $\cGamma$ encodes information about all the correlations in the theory. Here, we define  $\cGamma$ and provide some of its properties.

For simplicity, we consider a field theory for a scalar field $\hphi(\x)$ in $d$ dimensions and we constrain the total magnetization $\hphi_0\equiv L^{-d}\int_\x \hphi(\x)$ to be given by a fixed value $s$. Here $\int_\x\equiv \int d^d x$ and $L^d$ is the volume of the $d$-dimensional system of finite size $L$ with periodic boundary conditions. The field is sampled with a microscopic action (or Hamiltonian) $\mathcal H[\hphi]$. We denote averages with respect to the Boltzmann weight, without the constraint, as $\langle \mathcal{O} \rangle\propto \int \DD[ \hphi] \mathcal{O} e^{-\mathcal H(\hphi)} $.  We assume that the theory is regularized at a microscopic scale $a$, e.g.  the lattice spacing of the underlying microscopic model.

The PDF $P_L(s)$ of the order parameter can be expressed as a constrained average,
\begin{equation}
   P_L(s)= e^{-L^d I(s)}\propto\int \DD \hphi \, \delta(\hphi_0-s) \, e^{-\mathcal H[\hphi]}. \label{eq:PDFdef}
\end{equation}
\Cref{eq:PDFdef}  defines the rate function $I(s)$, a quantity known in quantum field theory as the constraint effective potential~\cite{Fukuda1975, ORaifeartaigh1986,Gockeler1991}.

Our goal is to define a generalization of the constraint effective potential, the constraint effective action, just as the effective action is of the effective potential. In field theory, the effective action $\Gamma[\phi]$, or Gibbs free energy,  is defined as the Legendre transform of the free energy $\log \mathcal Z[h]/\mathcal Z[0]=\log \langle e^{h.\hphi}\rangle$, with $h.\hphi=\int_\x h(\x)\hphi(\x)$, i.e. $\Gamma[\phi]=-\ln \mathcal Z[h]+h.\phi$ with $h(\x)=\frac{\delta \Gamma}{\delta \phi(\x)}$. Naively, one could hope to define $I(s)$ as the Legendre transform of the logarithm of $\check{\mathcal{Z}}[h]=\langle  e^{h.\hphi}\delta(\hphi_0-s)\rangle$. However, since in constant field $h(\x)=h$, $\check{\mathcal{Z}}[h]\propto e^{h L^d s} P_L(s)$, its logarithm is linear in $h$ and the Legendre transform is ill-defined.

To remediate this problem, we follow \cite{Balog2022} and mollify the delta constraint as a Gaussian with parameter $M$,
\begin{equation}
\mathcal Z_{M}[h] =\int \DD\hphi\, e^{-\mathcal H_M[\hphi] + h.\hphi },
\end{equation}
where ${\cal H}_M[\hphi]={\cal H}[\hphi]+\frac {M^2}2 [\int_\x (\hphi(\x)-s)]^2$. Here, the $M$-dependent part plays the role of a soft constraint on the zero-momentum mode of the field. In the limit $M\to\infty$, this constraint becomes, up to a constant, a Dirac-delta, and we get $\lim_{M\to\infty }\mathcal Z_{M}[h] \propto \check{\mathcal{Z}}[h]$. Similarly, we note the limit $M\to\infty$ of any quantity $O_M$ by $\check O$. In the opposite limit, we recover the standard generating function $\lim_{M\to 0 }\mathcal Z_{M}[h] = {\mathcal{Z}}[h]$.

The soft constraint makes it  possible to define a modified Legendre transform, using  $\phi(\x)=\frac{\delta \ln{ \mathcal Z}_{M}}{\delta h(\x)}$, as
\begin{equation}
\Gamma_{M}[\phi]=-\ln \mathcal Z_{M}[h]+h.\phi-\frac {M^2}2 \left[\int_\x (\phi(\x)-s)\right]^2,
\label{eq_def_GammaA_LT}
\end{equation}
with $h(\x)$ given by (and thus depending implicitly on $\phi(\x)$ and $s$)
\begin{equation}
\frac{\delta\Gamma_{M}}{\delta\phi(\x)}=h(\x)-M^2\int_\y(\phi(\y)-s).
\end{equation}
This in turn implies
\begin{equation}
\begin{split}
e^{-\Gamma_{M}[\phi]}=\int \DD\hat\phi\, e^{-{\cal H}(\hphi)+\frac{\delta\Gamma_{M}}{\delta\phi}.(\hphi-\phi) } e^{-\frac {M^2}2 \left[\int_\x (\hphi(\x)-\phi(\x))\right]^2}.
\end{split}
\label{eq:expGammaM}
\end{equation}
Note that due to the definition of the modified Legendre transform, $\Gamma_{M}[\phi]$ does not depend on $s$, even though $\mathcal Z_{M}[h]$ does so implicitly.

Let us now give some physical interpretation to $\Gamma_{M}$. First, for $M=0$, one recovers the standard effective action $\lim_{M\to 0 }\Gamma_{M}[\phi] = \Gamma[\phi]$. In the opposite limit, we obtained the constraint effective action $\cGamma[\phi] \equiv \lim_{M\to\infty}\Gamma_{M}[\phi] $.
When evaluated in a constant field $\phi(\x)=\bar\phi$,
\begin{equation}
\frac{\delta\Gamma_{M}}{\delta\phi(\x)}\bigg|_{\phi(\x)=\bar\phi}=h-M^2L^d(\bar\phi-s)
\end{equation}
 is a constant, such that $\langle\hat\phi(\x)\rangle =\bar\phi$ by translation invariance. Therefore, 
\begin{equation}
e^{-\Gamma_{M}[\phi(\x)=\bar\phi]}=\int \DD\hat\phi\, e^{-{\cal H}(\hat\phi) -\frac {M^2}2[\int_\x(\hat\phi(\x)-\bar\phi)]^2 +\int_\x(\hat\phi(\x)-\bar\phi)\frac{\delta\Gamma_{M}}{\delta\phi(\x)}\big|_{\bar\phi}},
\end{equation}
and thus
\begin{equation}
\begin{split}
    \lim_{M\to\infty} e^{-\Gamma_{M}[\phi(\x)=\bar\phi]}&\propto\int \DD\hat\phi\, \delta\left(\bar\phi-\hphi_0\right)e^{-{\cal H}(\hat\phi)}\\
& \propto P_L(s=\bar\phi),
\end{split}
\label{eq:P}
\end{equation}    
that is,  $I(s=\bar\phi)=L^{-d}\cGamma[\phi=\bar\phi]$ just as, in the absence of constraint, the effective potential is given by $U(\bar\phi)=L^{-d}\Gamma[\phi=\bar\phi]$.
As $\bar\phi$ and $s$ play the same role in \cref{eq:P}, we will use $s$ instead of $\bar\phi$ in the remainder of the manuscript.
We further note that $\cGamma[\phi=s]$ is not the Legendre transform of $\ln\check{\mathcal{Z}}[h]$ in constant field, since the $M\to \infty$  limit and the modified Legendre transform do not commute. Indeed $\cGamma[\phi]$ is well defined for all $\phi$ because we subtracted an infinite term in \cref{eq_def_GammaA_LT}.

The interpretation of the constraint effective action for a non-constant field is less obvious. However, its functional derivatives, evaluated in a constant field $\phi(\x)=s$ have a clear meaning. We introduce the $n$-point connected correlation functions 
\begin{equation}
\begin{split}
    G_M^{(n)}(\{\x_i\}_{i=1,\ldots,n};s)&= \frac{\delta^n  \ln\mathcal Z_M[h]}{\delta h(\x_1)\ldots \delta h(\x_n)}\bigg|_{h(\x)=\bar h},
\end{split}
\end{equation}
such that $\check G^{(n)}(\{\x_i\}_{i=1,\ldots,n};s)$ is the connected $n$-point correlation function at fixed magnetization. For instance, for $n=2$, we have
\begin{equation}
\begin{split}
   \check G(\x_1,\x_2;s)\equiv \check G^{(2)}(\x_1,\x_2;s)&= \frac{\langle \hphi(\x_1)\hphi(\x_2) \delta(s-\hphi_0)\rangle}{\langle  \delta(s-\hphi_0)\rangle}-\frac{\langle \hphi(\x_1) \delta(s-\hphi_0)\rangle}{\langle  \delta(s-\hphi_0)\rangle}\frac{\langle \hphi(\x_2) \delta(s-\hphi_0)\rangle}{\langle  \delta(s-\hphi_0)\rangle}.
\end{split}
\label{eq:G2c}
\end{equation}

It is well known that the second functional derivative of the \emph{standard} effective action is the inverse of the two-point correlation function. For the constrained effective action, the result almost holds, except at vanishing momentum. Indeed, standard manipulations involving Legendre transforms, adapted to the modified definition used here, show that 
\begin{equation}
    \int_\y \left(\frac{\delta^2 \Gamma_M[\phi]}{\delta\phi(\x)\delta\phi(\y)}+M^2\right)\frac{\delta^2\ln\mathcal Z_M}{\delta h(\y)\delta h(\z)}=\delta(\x-\z).
\end{equation}
Evaluating this result in a constant field $\phi(\x)=s$, and calling the $n$-point vertex function in constant field 
\begin{equation}
\begin{split}
    \Gamma^{(n)}_M(\{\x_i\}_{i=1,\ldots,n};s)&= \frac{\delta^n  \Gamma_M[\phi]}{\delta \phi(\x_1)\ldots \delta \phi(\x_n)}\bigg|_{\phi(\x)=s},
\end{split}
\end{equation}
we obtain
\begin{equation}
    \int_\y \left(\Gamma^{(2)}_M(\x,\y;s)+M^2\right)G_M(\y,\z;s)=\delta(\x-\z),
\end{equation}
or, using translation invariance and going to momentum space, 
\begin{equation}
   G_M(\p;s)\equiv G^{(2)}_M(\p,-\p;s)  =\left(\Gamma^{(2)}_M(\p;s)+M^2\delta_{\p,\zero}\right)^{-1}.
   \label{eq:G_M}
\end{equation}
In a finite-size system the momenta are discrete, $\p = 2\pi \n /L$, $\n\in \mathbb{Z}^d$, and $\delta_{\p,\zero}$ stands for the Kronecker delta.
In particular, since $\lim_{M\to\infty }\Gamma^{(2)}_M=\cGamma^{(2)}$ is well defined, we get that
\begin{equation}
   \check G(\p;s)=
 \begin{cases}
      0 & \text{if $\p=\zero$},\\
      \cGamma^{(2)}(\p;s)^{-1} & \text{otherwise}.
    \end{cases}   
\end{equation}
The fact that $\check G(\p=0;s)=0$ is a sum-rule implied by the constraint, since from \cref{eq:G2c}
\begin{equation}
\begin{split}
    \check G(\p=\zero;s)&\propto L^{-d}\int_\x \check G(\x,\y;s)\\
                    &=\frac{\langle \hphi_0\hphi(\y) \delta(s-\hphi_0)\rangle}{\langle \delta(s-\hphi_0)\rangle}-s^2\\
                    &=0,
\end{split}
\end{equation}
where we used $\langle \hphi(\y) \delta(s-\hphi_0)\rangle=s \langle \delta(s-\hphi_0)\rangle$ by translation invariance. Importantly, in our framework the sum-rule comes from the $M^2\to\infty$ term in the right-hand side of \cref{eq:G_M} and not from the behavior of $\cGamma^{(2)}(\p=\zero;s)$. In fact, using the method outlined in \cite{Blaizot2006} one readily shows that $\cGamma^{(2)}(\zero=0;s)= I''(s)$ (where the primes denote derivatives with respect to the argument). As we will see in \cref{sec:finite_size}, the behavior of $\cGamma^{(2)}(\p;s)$ at $\p\neq \zero$ and that of $\cGamma^{(2)}(\p=\zero;s)$ are in a certain sense unrelated.

The relationship between arbitrary $n$-point correlation functions and vertex functions follows from textbook methods~\cite{DupuisBook1}, while the sum-rule also generalizes to $n$-point correlation functions when any one momentum vanishes. For instance,
\begin{equation}
\check G^{(3)}(\x,\y,\z;s)=-\int_{\x',\y',\z'} \cGamma^{(3)}(\x',\y',\z';s)\check G(\x',\x;s)\check G(\y',\y;s)\check G(\z',\z;s),
\end{equation}
and $\int_\x \check G^{(3)}(\x,\y,\z;s)=\int_\y \check G^{(3)}(\x,\y,\z;s)=\int_\z \check G^{(3)}(\x,\y,\z;s)=0$.

Of course, computing $\cGamma$ is as hard as computing any generating function and cannot be done exactly in general. In the following, we leverage the methodology of FRG to compute approximate flows of the constraint effective action.

\section{Scale-dependent constraint effective action \label{sec:FRG}}

\subsection{Definition and exact flow equation}

To implement the machinery of the FRG, still following \cite{Balog2022}, we construct a one-parameter family of models denoted by ${\cal Z}_{M,k}[h]$ by modifying the original Hamiltonian ${\cal H}_M$ into ${\cal H}_M + \Delta {\cal H}_k $.
Here, $\Delta {\cal H}_k$ is a term aimed at effectively freezing the low wavenumber fluctuations $\hat\phi(|\mathbf{q}| < k)$ while leaving the high wavenumber modes $\hat\phi(|\mathbf{q}| > k)$ unchanged. It is chosen to be quadratic, $\Delta {\cal H}_k=1/2\, \hat \phi.R_k.\hat \phi$,
where $R_k(\mathbf{x}, \mathbf{y})$ satisfies the following conditions: (i) when $k \sim a^{-1}$, $R_{k \sim a^{-1}}(|\mathbf{q}|)$ is very large for all $|\mathbf{q}|$, implying that all fluctuations are frozen; and (ii) $R_{k=0}(|\mathbf{q}|) \equiv 0$, so that all fluctuations are integrated over, and ${\cal Z}_{M,k=0}[h] = {\cal Z}_M[h]$. Varying the scale $k$ between $a^{-1}$ and 0 induces the RG flow of ${\cal Z}_{M,k}[h]$, where fluctuations of wavenumbers $|\mathbf{q}| > k$ are progressively integrated over.

Defining the scale-dependent constraint effective action as
\begin{equation}
\Gamma_{M,k}[\phi]=-\ln \mathcal Z_{M,k}[h]+h.\phi-\frac12\phi. R_k.\phi-\frac {M^2}2 \left[\int_\x (\phi(\x)-s)\right]^2,
\label{eq:def_GammaA_LT}
\end{equation}
the equivalent of \cref{eq:expGammaM} is
\begin{equation}
e^{-\Gamma_{M,k}[\phi]}=\int \DD\hphi\, e^{-{\cal H}[\hphi] -\frac12 (\hat\phi-\phi).R_{M,k}.(\hat\phi-\phi) +\frac{\delta\Gamma_{M,k}}{\delta\phi}.(\hat\phi-\phi)},
\label{eq:second-def-gammak}
\end{equation}
where $R_{M,k}(\x,\y)=R_k(\x,\y)+M^2$, or in momentum space $R_{M,k}(\q)=R_k(\q)+M^2\delta_{\q,\zero}$. Up to the $M^2$ term in $R_{M,k}$, \cref{eq:second-def-gammak} is formally identical to the usual scale-dependent  effective action $\Gamma_k$ introduced in FRG \cite{Dupuis2021}, and in particular $\Gamma_{k}[\phi]=\Gamma_{M=0,k}[\phi]$. The exact FRG equation satisfied by $\Gamma_{M,k}[\phi]$ is the usual Wetterich equation in the presence of the regulator $R_{M,k}$:
\begin{equation}
\partial_k \Gamma_{M,k}[\phi]=\frac{1}{2}\int_{\x,\y}\partial_k R_{M,k}(\x,\y)\left(\Gamma_{M,k}^{(2)}+R_{M,k}\right)^{-1}(\x,\y),
\label{eq:Wetterich}
\end{equation} 
where $\Gamma_{M,k}^{(2)}=\Gamma_{M,k}^{(2)}[\x,\y;\phi]=\frac{\delta^2\Gamma_{M,k} }{\delta\phi(\x)\delta\phi(\y)}$. The flow equation of $\cGamma_k=\lim_{M\to\infty }\Gamma_{M,k}$ is given by the limit $M\to \infty $ of the right-hand-side of \cref{eq:Wetterich}, i.e. the derivative $\partial_k$ and the limit $M\to\infty$ commute.
This equation has been solved for the three-dimensional Ising universality class in the simplest approximation, the local potential approximation (LPA) in \cite{Balog2022} and in \cite{rancon_probability_2025} for the O($N$) model, with good agreement for the shape of the rate function between FRG and Monte Carlo simulations.

Defining as above the $n$-point (scale-dependent) vertex function in a constant field 
\begin{equation}
\begin{split}
    \Gamma^{(n)}_{M,k}(\{\x_i\}_{i=1,\ldots,n};s)&= \frac{\delta^n  \Gamma_{M,k}[\phi]}{\delta \phi(\x_1)\ldots \delta \phi(\x_n)}\bigg|_{\phi(\x)=s},
\end{split}
\end{equation}
as well as the scale-dependent effective potential $U_{M,k}(s) = L^{-d}\Gamma_{M,k}[\phi]\big|_{\phi(\x)=s}$, and using translational invariance when the field is constant, we obtain a hierarchy of flow equations
\begin{align}
\partial_k U_{M,k}(s)={}&\frac{1}{2L^d}\sum_{\q}\partial_k R_{M,k}(\q)G_{M,k}(\q;s),
\label{eq:flow_vertexU}
\\
\partial_k \Gamma^{(2)}_{M,k}(\p;s) ={}& -\frac{1}{2L^d}\sum_{\q}\partial_k R_{M,k}(q)G_{M,k}^2(\q;s)\Big( \Gamma^{(4)}_{M,k}(\p,-\p,\q,-\q;s)\nnl
&-2\Gamma^{(3)}_{M,k}(\p,\q,-\q-\p;s)G_{M,k}(\q+\p;s) \Gamma^{(3)}_{M,k}(-\p,-\q,\p+\q;s) \Big),
\label{eq:flow_vertexG2}
\end{align} 
where $G_{M,k}(\q;s)=\left(\Gamma_{M,k}^{(2)}(\q;s)+R_{M,k}(\q)\right)^{-1}$. 

The corresponding equations for the scale-dependent constraint action are obtained by taking the limit $M\to\infty$. As discussed in  \cref{sec:const}, the vertex functions are well behaved in this limit (the inclusion of the regulator does not change this), $\lim_{M\to\infty}U_{M,k}(s)=\cU_k(s)$ and $\lim_{M\to\infty}\Gamma^{(n)}_{M,k}=\cGamma^{(n)}_k$, while
\begin{equation}
  \lim_{M\to\infty} G_{M,k}(\q;s)=\check G_k(\q;s)=
 \begin{cases}
      0 & \text{if $\q=\zero$},\\
      \left(\cGamma^{(2)}(\q;s)+R_k(\q)\right)^{-1} & \text{otherwise}.
    \end{cases}   
\end{equation}
Noting further  that $\partial_k R_{M,k} = \partial_k R_{k}$, the flow equations \cref{eq:flow_vertexU,eq:flow_vertexG2} become
\begin{align}
\label{eq:flowcU}
\partial_k \cU_{k}(s)&=\frac{1}{2L^d}\sum_{\q\neq 0}\partial_k R_{k}(\q)\cG_{k}(\q;s),\\
\label{eq:flow_2vertex_check}
\partial_k \cGamma^{(2)}_{k}(\p;s) ={}& \frac{1}{2L^d}\sum_{\q\neq\zero}\partial_k\cG_{k}(\q;s)\cGamma^{(4)}_{k}(\p,-\p,\q,-\q;s)\nonumber\\
&-\frac{1}{2L^d}\sum_{\q\neq\zero,-\p}\partial_k\left(\cG_{k}(\q;s)\cG_{k}(\q+\p;s)\right) \cGamma^{(3)}_{k}(\p,\q,-\q-\p;s)\cGamma^{(3)}_{k}(-\p,-\q,\p+\q;s).
\end{align} 
Note that these two equations are consistent with the fact that $\cGamma^{(2)}_{k}(\zero;s)=\cU_{k}''(s)$. Indeed, for $\p=\zero$, the second reads
\begin{equation}
\label{eq:gamm20}
\begin{split}
\partial_k \cGamma^{(2)}_{k}(\zero;s) &= \frac{1}{2L^d}\sum_{\q\neq\zero}\partial_k\cG_{k}(\q;s)\cGamma^{(4)}_{k}(\zero,\zero,\q,-\q;s)-\frac{1}{2L^d}\sum_{\q\neq\zero}\partial_k\left(\cG^2_{k}(\q;s)\right) \left(\cGamma^{(3)}_{k}(\zero,\q,-\q;s)\right)^2\\
 &= \frac{1}{2L^d}\sum_{\q\neq0}\partial_k\cG_{k}(\q;s)\partial_s^2\cGamma^{(2)}_{k}(\q;s)-\frac{1}{2L^d}\sum_{\q\neq\zero}\partial_k\left(\cG^2_{k}(\q;s)\right) \left(\partial_s\cGamma^{(2)}_{k}(\q,-\q;s)\right)^2\\
 &=\partial_s^2\left(\frac{1}{2L^d}\sum_{\q\neq \zero}\partial_k R_{k}(\q)\cG_{k}(\q;s)\right)\\
 &=\partial_k \cU_{k}''(s),
\end{split}
\end{equation} 
where we have used the exact relationship between vertex functions \cite{Blaizot2006}
\begin{equation}
    \begin{split}
        \cGamma^{(3)}_k(\p,-\p,\zero;s)=\partial_s\cGamma^{(2)}_k(\p;s),\\
        \cGamma^{(4)}_k(\p,-\p,\zero,\zero;s)=\partial_s^2\cGamma^{(2)}_k(\p;s).
    \end{split}
    \label{eq:relation_vertex}
\end{equation}

\begin{figure}[t!]
    \centering
    \null\hfill
    \includegraphics[scale=1]{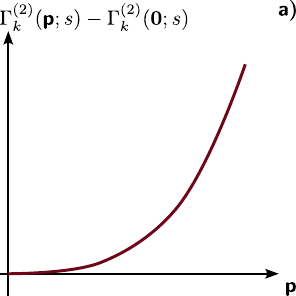}
    \hfill
    \includegraphics[scale=1]{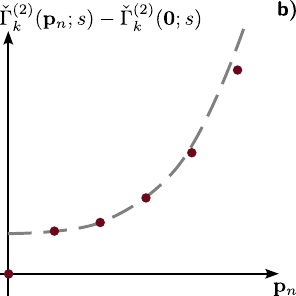}
    \hfill\null
    \caption{
    Sketches illustrating the small-momentum behavior of the vertices $\Gamma_k^{(2)}$ (in the thermodynamic limit) and $\cGamma_k^{(2)}$ (in a constrained finite-size system). In the thermodynamic limit [panel a)] $\Gamma_k^{(2)}(\pc;s)-\Gamma_k^{(2)}(\zero;s)$, represented by the red curve, vanishes quadratically. For the finite-size system [panel b)], the allowed momenta are quantized and the values of $\cGamma_k^{(2)}(\p_n;s)-\cGamma_k^{(2)}(\zero;s)$ are represented with red dots. Here $\p_n=2\pi n/L(1,0,\dots, 0)$, $n=0,1,2,\dotsc$. For $\p_n=\zero$, $\cGamma_k^{(2)}(\p_n;s)-\cGamma_k^{(2)}(0;s)=0$ while for small $\p_n \neq \zero$ these values are well approximated by a parabola (dashed gray line) of equation $\cDelta_{2,k}(s)+\p_n^2 \cZ_k(s)$.
    }
    \label{fig:sketchFig}
\end{figure}

\subsection{Finite-size effect \label{sec:finite_size}}

At finite size, since the momenta $\p = 2\pi \n /L$, $\n\in \mathbb{Z}^d$ are quantized, the behavior of the vertex $\cGamma^{(2)}_k(\p;s)$ at small momenta has to be investigated with special care.
Physically, the reason why one expects $\cGamma^{(2)}_k(\zero;s)$ to behave differently from $\cGamma^{(2)}_k(\p;s)$ for $\p\neq \zero$ arises from the construction of the constraint effective action, with an infinite mass affecting only the zero-momentum mode. This is already evidenced in the properties of the constraint correlation function, which vanishes at zero momenta by the sum rule while for $\p\neq\zero$, $\cG(\p;s)^{-1}=\cGamma^{(2)}(\p;s)$ [see \cref{eq:G_M}].

This is reflected at the level of the flow equations. A comparison of \cref{eq:flow_2vertex_check,eq:gamm20} reveals a peculiar difference between the flow equation of $\cGamma^{(2)}_{k}(\zero;s)$ and $\cGamma^{(2)}_{k}(\p;s)$ for $\p \neq \zero$. Indeed, for $\cGamma^{(2)}_{k}(0;s)$ only the $\q=\zero$ term is absent from the sum in the bubble diagram (the second term of both \cref{eq:flow_2vertex_check,eq:gamm20}), while for $\cGamma^{(2)}_{k}(\p;s)$ both the $\q=\zero$ and $\q=-\p$ terms are absent. As a consequence, this discrepancy results in a peculiar behavior of the vertex at $\p=\zero$, see \cref{fig:sketchFig}. Indeed, while one can formally describe the $\p$ dependence of $\cGamma^{(2)}_{k}(\p;s)$ at small $\p$ as
\begin{equation}
\begin{split}
  \cGamma^{(2)}_{k}(\p;s)&= \cGamma^{(2)}_{k}(\zero;s)+(1-\delta_{\p,\zero})\cDelta_{2,k}(s)+\cZ_k(s)\p^2+O(|\p|^4),
\end{split}
  \label{eq:expansionSmp}
\end{equation}
the introduction of a nonzero function $\cDelta_{2,k}(s)$ becomes necessary.\footnote{This behavior has also been observed in a zero-dimensional quantum system, see \cite[Appendix B]{Rancon2014b}. In that context, the finite temperature implies that Matsubara frequencies are discrete, and some vertex functions have a non-trivial frequency dependence that prevents a naive interpolation to zero frequency. }  
In the $L\to\infty$ limit, $\cGamma^{(2)}_{k}(\p;s)$ becomes a function of a continuous variable with $\cDelta_{2,k}(s)\to 0$ and then \cref{eq:expansionSmp} reduces to the usual small momenta expansion of the inverse propagator. 
  In the thermodynamic limit, sums become integrals,
\begin{equation}
\frac{1}{L^d}\sum_\q f(\q)\to\int_{\qc}f(\qc)\equiv\int\frac{d^d \qc}{(2\pi)^d}f(\qc),
\end{equation}
which are insensitive to the absence of isolated points (such as $\q=\zero$ and $\q=-\p$).  Accordingly, $\partial_k \cDelta_{2,k}=O(L^{-d})$ vanishes so $\cDelta_{2,k}(s)=0$ while $\cZ_k(s)=\partial_{{\pc}^2}\cGamma^{(2)}_k(\pc;s)\big|_{\pc = 0}$ and \cref{eq:flow_2vertex_check} reduces to the FRG equation of the Ising model in the thermodynamic limit \cite{Benitez2012}.

The corresponding behavior for higher-order vertices $\cGamma^{(n)}_k$, which requires the introduction of a corresponding $\cDelta_{n,k}(s)$, is discussed in \cref{app:vertex}.

\subsection{Derivative expansion \label{sec:DE2}}

The most commonly used approximation scheme for FRG is the derivative expansion, where one assumes that the effective action can be expanded in the gradient of the field. 
Naively, the corresponding ansatz to second order (DE2) for the $\mathbb Z_2$ universality class would read
\begin{equation}
\cGamma_k[\phi]=\int_\x\left(\cZ_k(\phi(\x))\frac{(\nabla\phi(\x))^2}2+\cU_k(\phi(\x))+\mathcal O(\nabla^4)\right).
\label{eq:ansatzDE2}
\end{equation}
If, in addition, one imposes $\cZ_k(\phi(\x))=1$, then one recovers the local potential approximation used in \cite{Balog2022} and \cite{rancon_probability_2025}.
The inverse propagator in a constant field corresponding to the ansatz (\ref{eq:ansatzDE2}) is 
\begin{equation}
\cGamma^{(2)}_k(\p;s)=\cU''_k(s)+\cZ_k(s) \p^2 +\mathcal O (\p^4).
\end{equation}
While perfectly justified in the thermodynamic limit, this ansatz  misses the zero-momentum behavior at finite size discussed above (\cref{eq:expansionSmp}).
There is no ansatz for $\cGamma$ based on the derivative expansion that would reproduce such behavior.
To remedy the problem, we assume that the two-point vertex takes the form similar to \cref{eq:expansionSmp},
\begin{equation}
      \cGamma^{(2)}_{k}(\p;s)= \cU''_k(s) +(1-\delta_{\p,\zero})\cDelta_{2,k}(s)+\cZ_k(s)\p^2,
\label{eq:DE2}
\end{equation}
with the definitions 
\begin{eqnarray}
\label{eq:defcZ}
     \cZ_k(s)&=&\frac{\cGamma^{(2)}_k(\p_2;s)-\cGamma^{(2)}_k(\p_1;s)}{\p_2^2-\p_1^2},\\
\label{eq:defDelta}
    \cDelta_{2,k}(s)&=&\cGamma^{(2)}_k(\p_1;s)-\cU''_k(s)-\p_1^2\cZ_k(s),
\end{eqnarray}
and $\p_n=2n\pi/L(1,0,\ldots,0)$. Note that in the thermodynamic limit the definition of the function $\cZ_k$ in Eq. (\ref{eq:defcZ}), reduces to a derivative $\cZ_k(s)=\partial_{{\pc}^2}\cGamma^{(2)}_k(\pc;s)\big|_{\pc = \zero}$, similar to a usual definition of the field renormalization function in the derivative expansion.

To derive the flow equation for $\cZ_k$, one needs the three- and four-point vertices $\cGamma_k^{(3)}$ and $\cGamma_k^{(4)}$. As discussed in \cref{app:vertex}, these vertices generally involve distinct ``finite-momentum shifts'' $\cDelta_{3,k}$ and $\cDelta_{4,k}$, defined analogously to $\cDelta_{2,k}$ but being independent functions. As the flow of $\cDelta_{n,k}$ depends on $\cDelta_{n,k+1}$ and $\cDelta_{n,k+2}$, the derivative expansion does not close the hierarchy of equations. One therefore needs an additional approximation. One possibility would be to just neglect these terms and equate them to zero; however, we have observed that doing so leads to numerical instabilities when integrating the flow equations.
Instead, here we assume that we can approximate $\cDelta_{3,k}(s)\simeq \cDelta_{2,k}'(s)$ and $\cDelta_{4,k}(s)\simeq \cDelta_{2,k}''(s)$ in analogy with \cref{eq:relation_vertex}.
We leave for future work the detailed study of the various $\cDelta_{n,k}(s)$ functions and the range of validity of this additional approximation.
Thus, we assume the following form for the three-point and four-point vertex functions that contribute to the flow of $\cGamma^{(2)}_k$ (with $\p,\q,\p+\q\neq\zero$) ,
\begin{equation}
\begin{split}
   \cGamma^{(3)}_{k}(\p,\q,-\p-\q;s)&=\cU_k^{(3)}(s)+\cDelta_{2,k}'(s)+\cZ_k'(s)(\p^2+\q^2+\p.\q),\\
\cGamma^{(4)}_{k}(\p,-\p,\q,-\q;s)&=\cU_k^{(4)}(s)+\cDelta_{2,k}''(s)+\cZ_k''(s)(\p^2+\q^2).
\end{split}
\label{eq:vertexDE2}
\end{equation}

The terms involving $\cU_k$ and $\cZ_k$ correspond to the standard vertex functions found in the literature in the thermodynamic limit, as obtained from \cref{eq:ansatzDE2}.  At second order in momenta, the terms written here are the ones allowed by $\pi/2$ rotations, parity, and exchange of momenta. While they are invariant under continuous rotations, additional terms are allowed at higher orders (e.g. $p_x^4+p_y^4+\ldots$) due to the breaking of that symmetry induced by the boundaries. 
For the constraint effective action, the DE2 approximation amounts to solving the coupled flows of $I_k$, $\cZ_k$ and $\cDelta_{2,k}$, contrary to the standard derivative expansion where only two functions are needed.

\section{Critical constraint effective action \label{sec:critical}}

\subsection{Scaling form}

We now study the properties of the constraint effective action at the critical point between the ferromagnetic and paramagnetic phases.
The corresponding observables are expected to be universal, however the universal scaling function depends on how the limit $T\to T_c$ and $L\to\infty$ are taken \cite{Balog2022}. For simplicity, we choose here to work directly at $T=T_c$ and consider sizes $L$ very large compared to the inverse UV cutoff (or lattice spacing).

At $T=T_c$, the rate function obeys the finite-size scaling law 
\begin{equation}
\label{eq:I_scaling}
I(s) = L^{-d}\tilde I(sL^{(d-2+\eta)/2}),
\end{equation}
where $\eta$ is the anomalous dimension ($\eta=1/4$ in two dimensions and $\eta\simeq 0.036$ in three dimensions), while
$\cGamma^{(2)}$ obeys the finite-size scaling law
\begin{equation}
\label{eq:gamm2_scaling}
\begin{split}
   \cGamma^{(2)}(\p;s) &= L^{-2+\eta}\tilde \Gamma(\p L; sL^{(d-2+\eta)/2}).
\end{split}
\end{equation}

Considering first the mean-field scaling $\Gamma^{(2)}(\p)-\Gamma^{(2)}(\zero)=\p^2$ and second the zero-mode discrepancy, it is convenient to work with the function ($n\geq 2$)
\begin{equation}
\label{eq:A_definition}
  A(\p_n;s) = \frac{\cGamma^{(2)}(\p_n;s)-\cGamma^{(2)}(\p_1;s)}{\p_n^2-\p^2_1},
\end{equation}
where $\p_n=2\pi n/L(1,0,\ldots,0)$. With this definition, $\cZ(s)=A(\p_2;s)$, see \cref{sec:DE2}.
Note that at criticality
\begin{equation}
\label{eq:A_scaling}
A(\p_n;s) = L^{\eta}\tilde A(\p_n L; sL^{(d-2+\eta)/2}).
\end{equation}

To compare observables of different models belonging to the same universality class, e.g. that of Ising lattice model to those of a continuum scalar field theory, one must account for two independent non-universal amplitudes: one fixing the order-parameter scale and one fixing the correlation‐length amplitude. 
This is the essence of two‐scale‐factor universality \cite{bervillier_universal_1976}. Furthermore, by a similar argument to that of Privman and Fisher \cite{privman_universal_1984}, one shows that  $L^d I$ is a universal function of the correctly rescaled magnetization, without any additional non-universal amplitude.\footnote{Privman and Fisher have argued that the singular part of the free energy density $f$ scales as $f(t,h)=L^{-d}\tilde f(L/\xi_0t^{\nu},L^{(d+2-\eta)/2}h/h_0)$, with $t=T/T_c-1$ and $h$ the magnetic field. The scaling function $\tilde f$ is universal when both non-universal amplitudes $\xi_0$ and $h_0$ are fixed (for instance, one can choose $\xi_0$ such that $\xi_0 t^{-\nu}$ equals the thermodynamic limit correlation length). The magnetization in turn obeys the scaling form $m=L^{-\frac{d-2+\eta}{2}}/h_0 \tilde f^{(0,1)}(L/\xi_0t^{-\nu},L^{\frac{d+2-\eta}{2}}h/h_0)$ with $\tilde f^{(0,1)}(x,y)=\partial_y \tilde f(x,y)$, and the Gibbs free energy density (related to the effective action in constant field) reads $g(t,m)=L^{-d} \tilde g(L/\xi_0t^{\nu},L^{(d-2+\eta)/2}mh_0)$ with $\tilde g$ the Legendre transform of $\tilde f$. It implies that the Gibbs free energy, and as a consequence the rate function, also obeys the two-scale universality. This corrects a wrong statement in \cite{Balog2022} which asserts that there is an additional amplitude associated with the scale of the rate function.} 
Here, as we focus on the case $T=T_c$, the correlation length in the thermodynamic limit is infinite and we do not need to fix the corresponding amplitude. See however \cite{Balog2022,sahu_generalization_2025,rancon_probability_2025} for discussions of this aspect in the context of the rate function.
To fix the scale of the magnetization $s_0$, we choose it to be the position of the minimum of the rate function, i.e. such that $I'(s_0)=0$. Note that $s_0= b\,L^{-(d-2+\eta)/2}$, where $b$ is the corresponding non-universal amplitude. Then all other quantities are measured in terms of $L$ and $s_0$, using the following rules
\begin{equation}
\begin{split}
    I(s) \to& L^d (I(s/s_0)-I(0)),\\
    A(\p_n;s)\to&  L^{d-2} s^2_0 A(\p_n L;s/s_0),\\
    \cZ(s) \to & L^{d-2} s^2_0 \cZ(s/s_0).
\label{eq:rescaling}
\end{split}
\end{equation}
We subtract a constant from the rate function, corresponding to the normalization of the PDF.
Note that $\cZ(s)$ scales as $L^{\eta}$, as expected for the field renormalization. All the results presented below have been rescaled following this procedure. Then the functions plotted are fully universal (i.e. completely independent of the microscopic details of the models).

\subsection{FRG results}

We have numerically integrated the flow equations of the functions $\cU_k$, $\cDelta_{2,k}$, and $\cZ_k$ at criticality in two and three dimensions, see \cref{app:FRG} for details. We now discuss the results and estimate the convergence, comparing our results to MC simulations in the next section.

\begin{figure}[t!]
    \centering
    \includegraphics[scale=0.5]{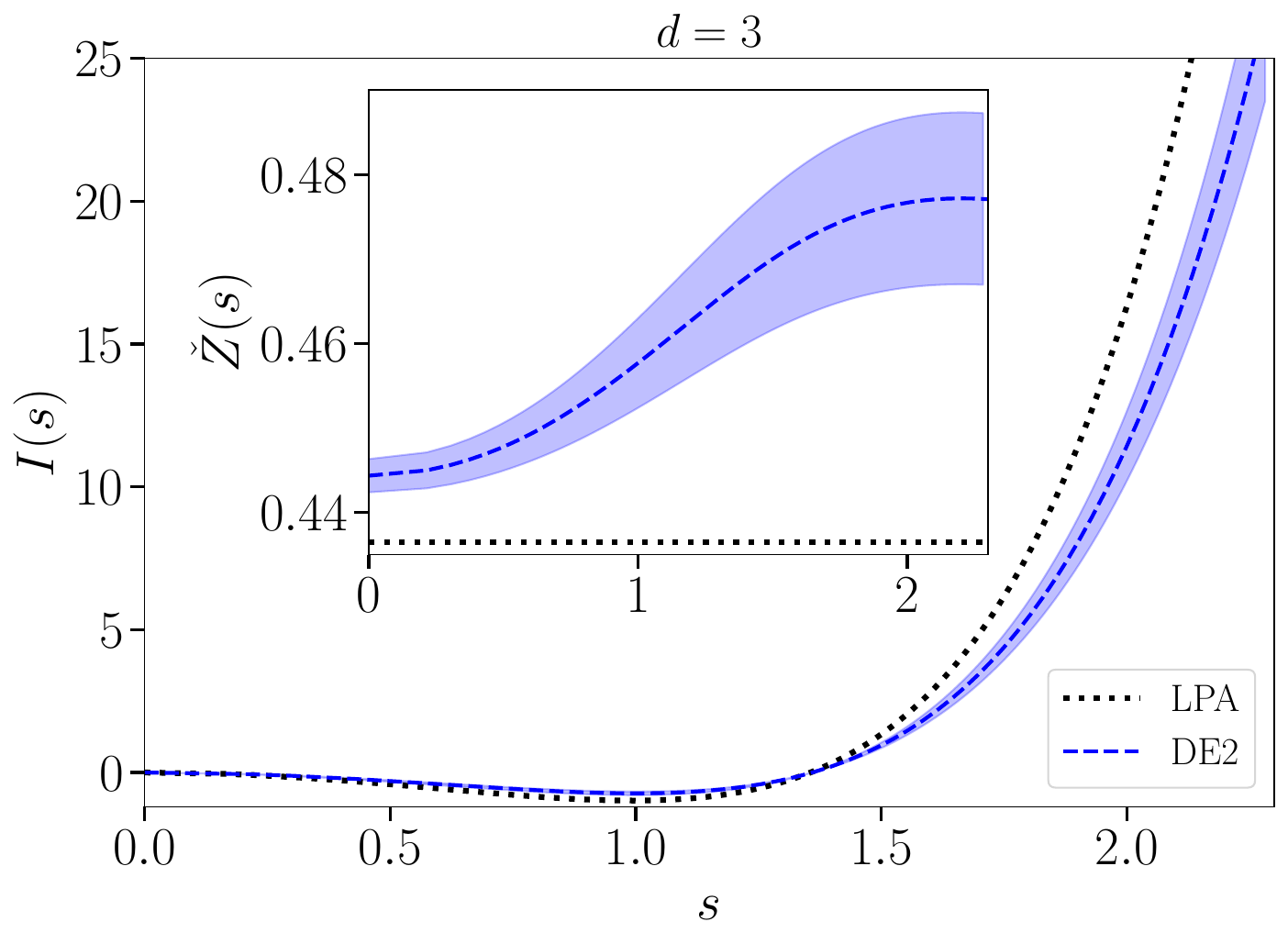}
    \caption{Main panel: critical rate function $I(s)$ from LPA and DE2 for Ising in three dimensions as a function of $s$. Inset: function $\cZ(s)$ as a function of $s$. The blue area corresponds to the error bar, see text. We used the optimal value of the regulator parameter $\alpha=4.65$  ($\alpha=1.3$) at LPA (DE2). }
    \label{fig:rate_3d}
\end{figure}

\begin{figure}[t!]
    \centering
    \includegraphics[scale=0.5]{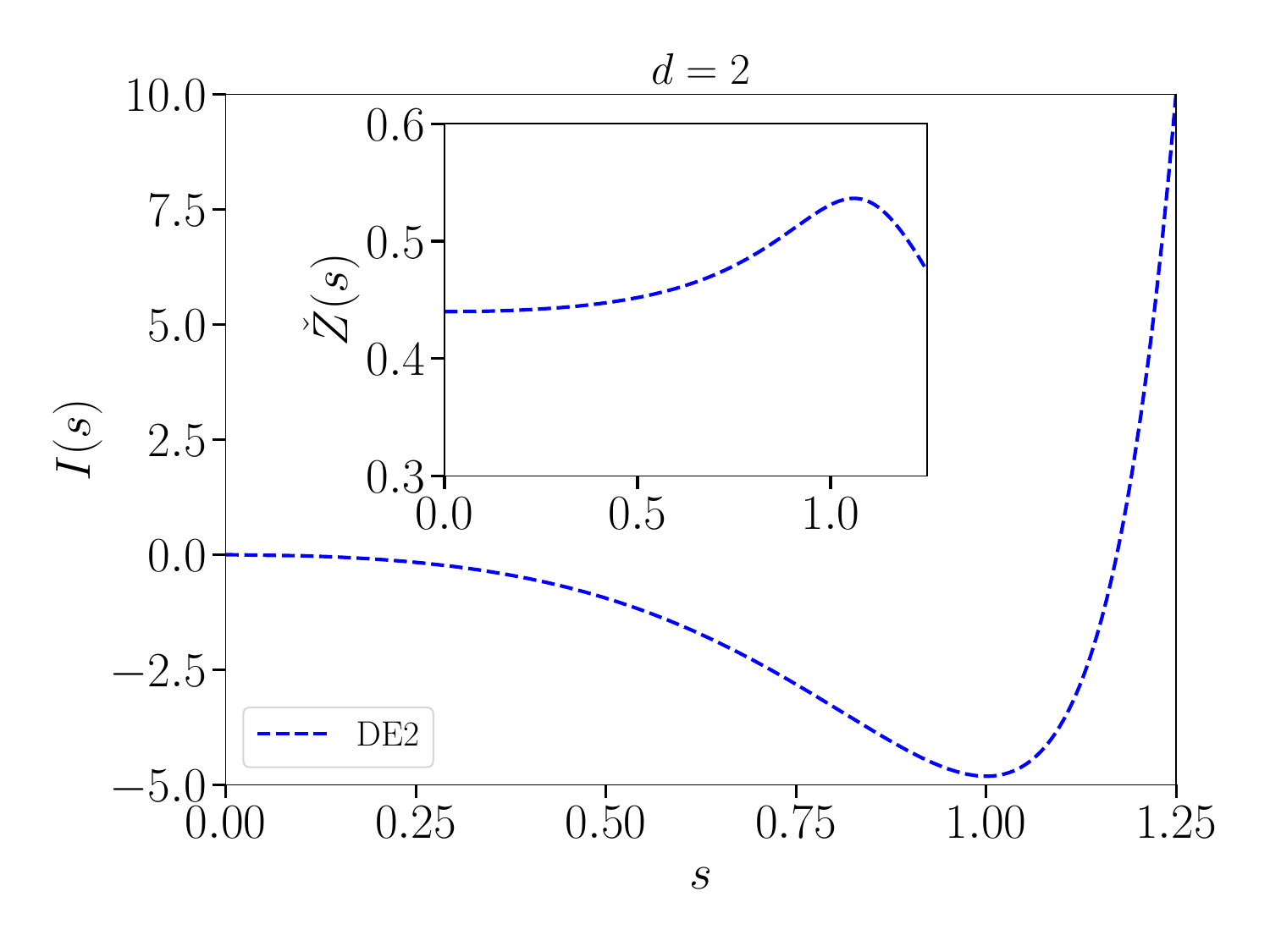}
    \caption{Main panel: critical rate function $I(s)$ obtained from DE2 in two dimensions as a function of $s$. Inset: function $\cZ(s)$ obtained at DE2. In both panels, the lines correspond to the optimal regulator parameter $\alpha=1$. The failure of LPA in two dimensions prevents the determination of error bars as in \cref{fig:rate_3d}, see text. }
    \label{fig:rate_2d}
\end{figure}

\cref{fig:rate_3d} and \cref{fig:rate_2d} show the rate functions $I(s)=\cU_{k=0}(s)$ and functions $\cZ(s)=\cZ_{k=0}(s)$ in dimensions three and two, respectively. We find that the rate function has a minimum and grows as a power law at larger $s$ (see also below). On the other hand, the function $\cZ$ has a rather weak variation, with a maximum slightly further than the minimum of the rate function.

\cref{fig:rate_3d} also shows the LPA results, which allow for computing the DE2 uncertainty (shown as shaded bands). To estimate this uncertainty, we compare two consecutive truncations—LPA and DE2—following the convergence arguments of \cite{Balog2019,de_polsi_precision_2020}.  In practice, each successive order in the derivative expansion reduces the error on observables by roughly a factor between $\tfrac{1}{9}$ and $\tfrac{1}{4}$.  We therefore define the most conservative error bars at DE2 order to be one‐quarter of the difference between the LPA and DE2 rate function curves.

All FRG computations in three dimensions use the exponential regulator (see Eq.~\eqref{eq:exp_reg}), with optimal prefactor $\alpha_{\rm opt}\approx4.65$ at LPA and $\alpha_{\rm opt}\approx1.3$ at DE2, obtained from the Principle of Minimal Sensitivity, see \cref{app:FRG}.  The dependence of our results on $\alpha$ is discussed in \cref{app:alpha}. At these parameters, the DE2 approximation yields $\eta\approx0.0455$ and $\nu\approx0.6275$, compared to the conformal‐bootstrap benchmarks $\eta_{\rm CB}=0.0363\ldots$ and $\nu_{\rm CB}=0.6300\ldots$~\cite{Kos2016}.  Varying the regulator shape has a negligible effect on the rate function relative to the LPA/DE2 difference. Note that this methodology gives us an error estimate for $\cZ$ as well, since, even though it is a constant in $s$ at LPA, it is still rescaled non-trivially using the prescriptions \cref{eq:rescaling}. 

Finally, the deterioration of DE2 accuracy in the far‐tail region ($s\gg1$) can be traced to the power law behavior in the tail, $I(s)\propto s^{\frac{2d}{d-2+\eta}}$ \cite{Bouchaud1990,Balog2022,balog_universal_2025}.
Since any truncated FRG scheme miscalculates $\eta$ (e.g.\ LPA predicts $\eta=0$), the tail’s exponent deviates from its exact value, and the error has to increase in this region. 

In two dimensions, the LPA truncation fails to find the phase transition, and hence cannot serve as a baseline for error estimation as it does in three dimensions.  A proper uncertainty analysis would require proceeding to the next order in the derivative expansion, which is beyond the scope of the present article.  Instead, we assess reliability by comparing critical exponents at DE2, using the exponential regulator with optimal prefactor $\alpha\approx1$, $\eta=0.293$ and $\nu=1.07$, to the exact ones, $\eta=\tfrac{1}{4}$ and $\nu=1$. We can thus estimate an error of about $10-20\%$.

\section{Comparison to Monte Carlo simulations \label{sec:MC}}

We have performed MC simulations of the two- and three-dimensional Ising model at $T=T_c$, for square (cubic) lattices with periodic boundary conditions of linear size $L=128,256,512$ ($16,32,64,128$). 
We limited ourselves to these system sizes to keep the volume of raw data manageable, particularly for the constrained correlation function. Extending the simulations to $L=1024$ in two dimensions or $L=256$ in three dimensions would substantially increase the computational and storage burden, despite yielding only marginal improvements in accuracy for corrections to scaling.

We used the Swendsen-Wang algorithm \cite{Swendsen1987} to sample spin configurations $\{\hat\sigma\}$ with the Boltzmann weight
\begin{equation}
\label{eq:BolzW}
  w(\{\hat\sigma\})\propto e^{-J/T \sum_{\langle \mathbf{i},\mathbf{j}\rangle}\hat\sigma_{\mathbf{i}}\hat\sigma_{\mathbf{j}}},
\end{equation}
where the sum runs over nearest neighbors.
The probability distribution function of the order parameter $P_L(s)$ is obtained by incrementing by one the corresponding bin of the PDF for each occurrence of $s$, where we used the method detailed in \cite{balog_universal_2025} to improve the statistics,  effectively corresponding  to about $10^{10}$ configurations for each size $L$.  The histogram is normalized at the end to determine the PDF.

\begin{figure}[t!]
    \centering
    \includegraphics[scale=0.5]{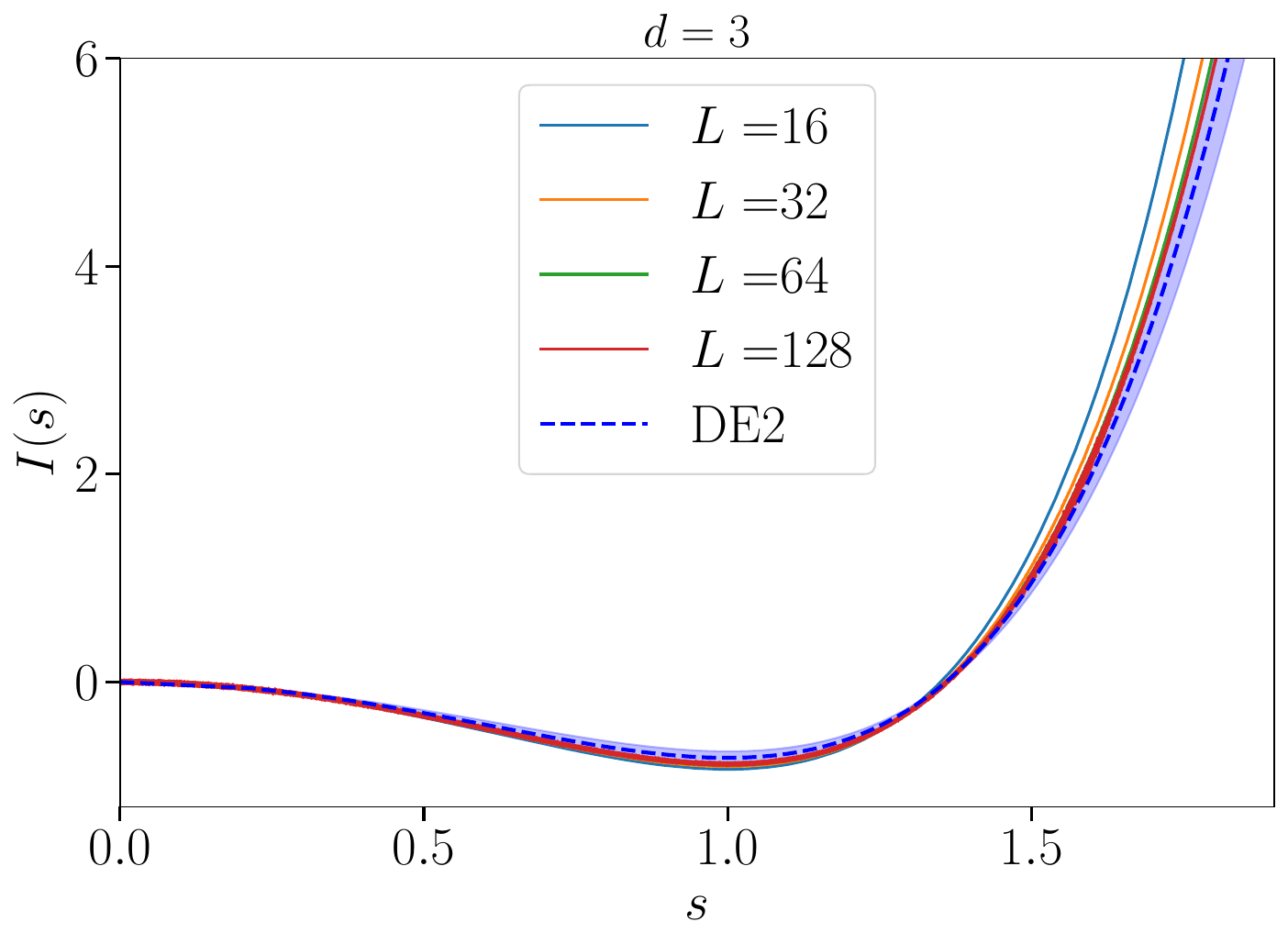}
    \caption{Critical rate function $I(s)$ of the Ising model in three dimensions from MC for different system sizes (full lines) and from the DE2 (dashed line). The shaded area gives the error bars for DE2, see text. }
    \label{fig:rate_3d_MC_FRG}
\end{figure}

\begin{figure}[t!]
    \centering
    \includegraphics[scale=0.5]{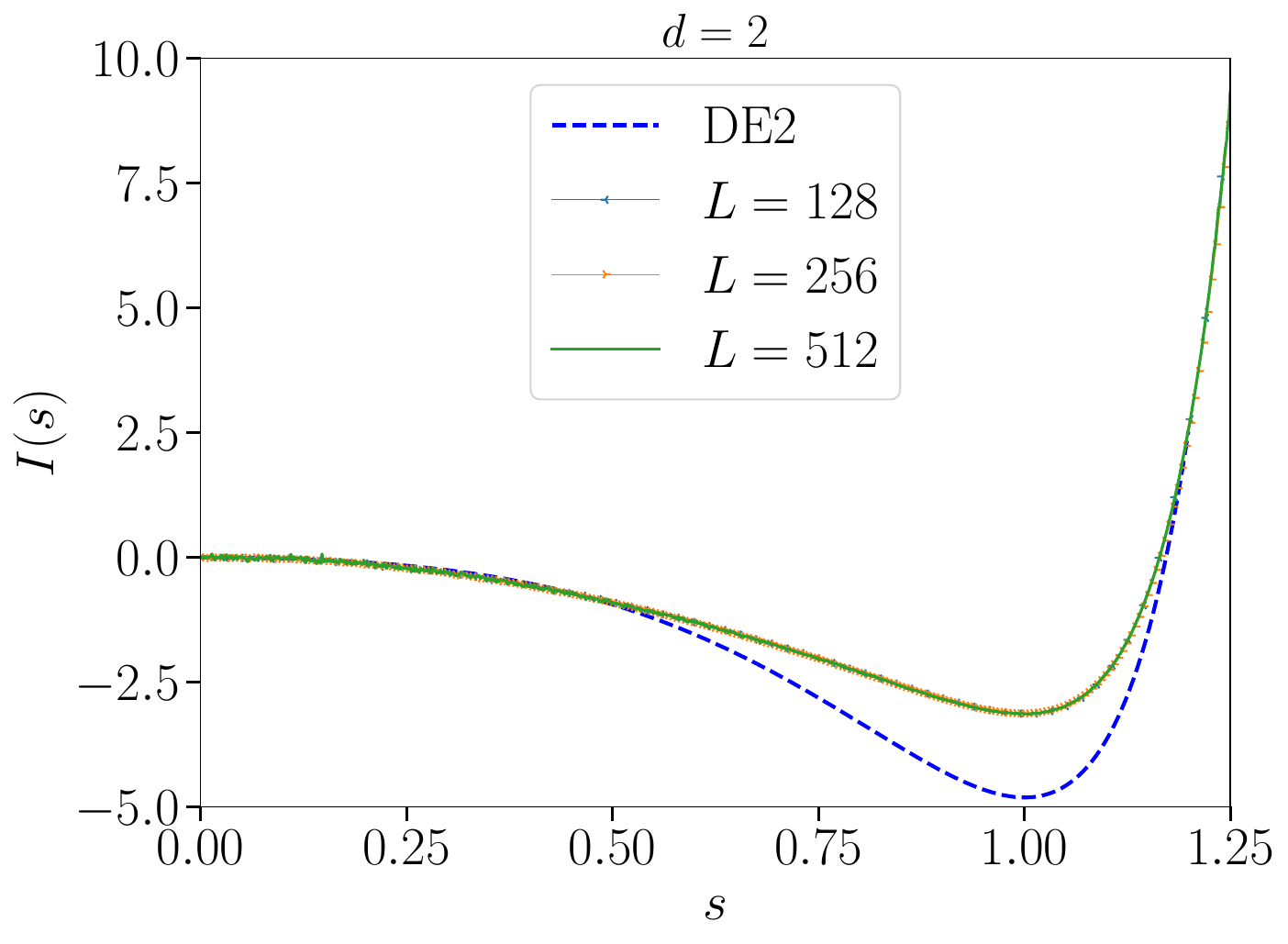}
    \caption{Critical rate function $I(s)$ of the Ising model in two dimensions from MC for different system sizes  (full lines) and DE2 (dashed line).}
    \label{fig:rate_2d_MC_FRG}
\end{figure}

For each occurrence of $s$, we also compute the correlation function in momentum space along the $x$-axis, e.g. $\cG(\p_n;s)$, $\p_n=2\pi n/L(1,0,\ldots,0)$, which  is obtained from the Fourier transform of 
\begin{equation}
 g(i;s)=L^{-2} \sum_{k,l}\langle\hat\sigma_{ (i,k)}\hat\sigma_{(0,l)}\delta(\sum_{\mathbf{j}}\hat\sigma_{\mathbf{j}} - L^2 s)\rangle
\end{equation}
in two dimensions (and similarly in three dimensions). In practice, to improve statistics, we compute the intermediate variable $\hat\Sigma_i=L^{-1}\sum_j \hat\sigma_{(i,j)}$, which corresponds to the average spin in the planes perpendicular to the $x$-axis at position $i$ and use the estimator $g(i;s)=L^{-1}\langle\sum_{j}\hat\Sigma_j \hat\Sigma_{i+j}\delta(\sum_{\mathbf{j}}\hat\sigma_{\mathbf{j}} - L^2 s)\rangle$.
This expression needs to be normalized by dividing it by $\langle\delta(\sum_{\mathbf{j}}\hat\sigma_{\mathbf{j}} - L^2 s)\rangle = P_L(s)$, giving $\check g(i;s)=g(i;s)/P_L(s)$, which obeys the sum-rule $L^{-1}\sum_i\check g(i;s)=s^2$. We have checked that this sum-rule is obeyed exactly in our simulations.
Finally, the constraint (connected) correlation function $\cG(\p_n;s)$ is given by
\begin{equation}
\label{eq:corrf_MC}
  \cG(\p_n;s) = \begin{cases}
      0 & \text{if $\p_n=\zero$},\\
      \frac{1}{L}\sum_j e^{i 2\pi nj/L}\check g(j;s)& \text{otherwise}.
  \end{cases}
\end{equation}

Running a numerical simulation with the Boltzmann weight \eqref{eq:BolzW}, one obtains sufficient sampling only for typical values of $s$, i.e, not too far from the minima of the rate function (the maxima of the PDF). This is especially relevant in two dimensions, since the rate function increases as $s^{16}$ already for $s\gtrsim s_0$. To increase the sampling of configurations with $s\gtrsim s_0$ in both dimensions, we have performed importance sampling by including a magnetic field to bias the PDF, and used multi-histogram reweighting to reconstruct the observables \cite{NewmanBarkema_book}.

\begin{table}[ht]
    \centering
    \begin{tabular}{|c|c|}
        \hline
        \multicolumn{2}{|c|}{\textbf{FRG}} \\ \hline
        LPA & $-0.983$ \\ 
        \hline
        DE2 & $-0.73(6)$ \\ 
         \hline
        \hline
        \multicolumn{2}{|c|}{\textbf{MC}} \\ \hline
        $L=16$ & $-0.8393(1)$ \\ \hline
        $L=32$ & $-0.8124(1)$ \\ \hline
        $L=64$ & $-0.7978(1)$ \\ \hline
        $L=128$ & $-0.7878(1)$ \\ \hline
    \end{tabular}
    \caption{Three-dimensional universal amplitude $\Delta I$ defined in \cref{eq:deltaI}, from FRG at LPA and DE2, and MC. For the estimation of error bars for DE2, see main text.}
    \label{tab:DeltaI}
\end{table}

We now compare the FRG results to MC simulations, starting with the rate function.
As shown in \cref{fig:rate_3d_MC_FRG}, in three dimensions, we obtain a very good agreement up to $s=1$. For larger values of $s$, we observe a discrepancy that we attribute to finite-size effects in the Monte Carlo simulations. Indeed, the agreement between MC and FRG improves as the system size is increased, and the former falls in the FRG error bars for $L>32$.

Table~\ref{tab:DeltaI} reports the universal amplitude
\begin{equation}
\label{eq:deltaI}
\Delta I=L^d(I(s_0)-I(0))
\end{equation} obtained from FRG and MC.  The LPA yields $\Delta I = -0.983$, whereas DE2 gives $\Delta I = -0.73(6)$.  The quoted uncertainty for DE2 corresponds to one quarter of the difference between the LPA and DE2 results, as explained above.  The MC results show $\Delta I$ moving from $-0.8393(1)$ toward $-0.7878(1)$ as $L$ increases, demonstrating convergence toward the thermodynamic limit value. This is consistent with the DE2 result within the error bars.

In two dimensions, \cref{fig:rate_2d_MC_FRG}, the rate function obtained from MC is already converged at $L=128$ for the range of $s$ accessible, and there are no visible finite-size corrections. We find that the DE2 overshoots quite strongly the value of the minimum of the rate function, with a $\Delta I\simeq -4.8$ compared to the MC result $\Delta I\simeq -3.13$. This shows that the DE2 is less reliable in two dimensions than in three dimensions, as expected. 

\begin{figure}
    \centering
    \includegraphics[scale=0.5]{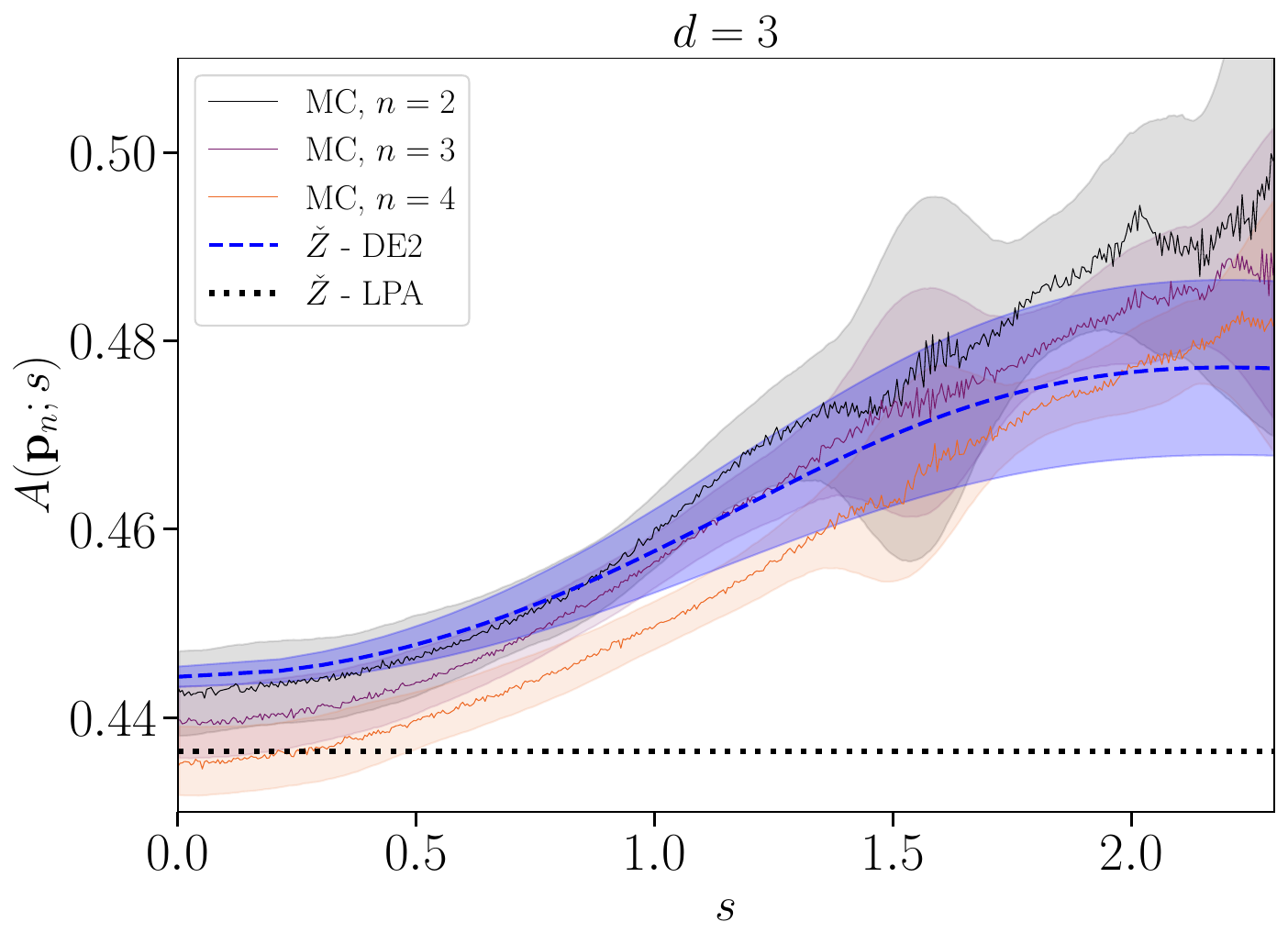}
    \caption{
    Function $A(\p_n;s)$ versus magnetization $s$ for various momenta in three dimensions. Solid curves show Monte Carlo data ($L=64$) at momenta $\p_n$. The data are noisy and are filtered for clarity using a cubic Savitzky–Golay filter over a hundred points, and is subsampled. The shaded bands indicate the data’s maximum and minimum fluctuations.  The dashed line is $\cZ(s)$, corresponding to $A(\p_2;s)$ at order DE2. The dotted line represents the results at LPA.  }
    \label{fig:Acphic_3d}
\end{figure}

\begin{figure}
    \centering
    \includegraphics[scale=0.5]{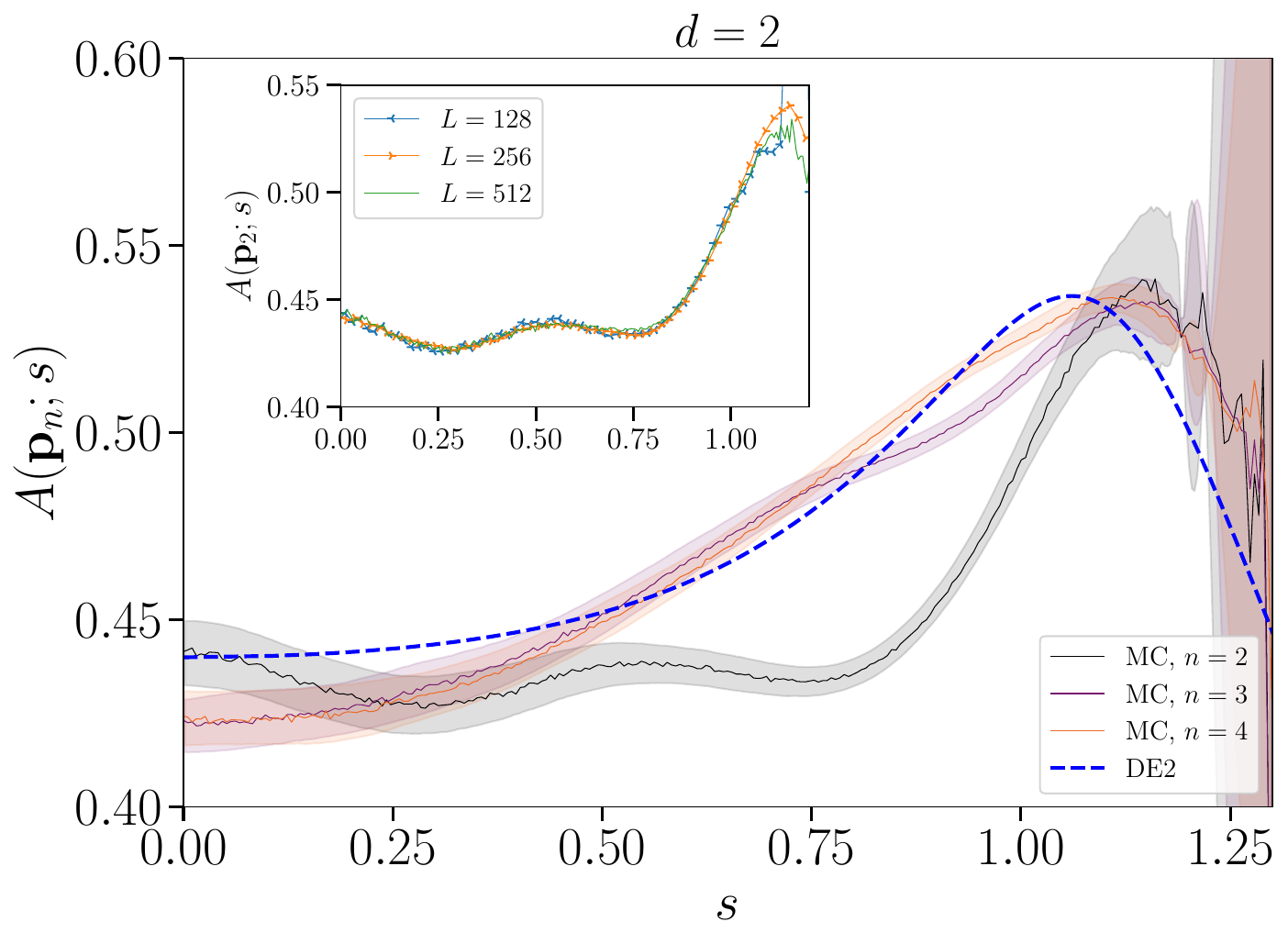}
    \caption{Function $A(\p_n;s)$ for several momenta in two dimensions, obtained from MC ($L=256$) and DE2. Inset: MC results for the function $A(\p_2;s)$ for various system sizes.  The results are converged and show scaling. Therefore, the oscillations in $s$ are not due to finite-size corrections. All MC data have been filtered for clarity using a cubic Savitzky–Golay filter over a hundred points, with shaded bands indicating the data’s maximum and minimum fluctuations.}
    \label{fig:Acphic}
\end{figure}

We now turn to the analysis of the momentum dependence of the constraint correlation functions, using the function $A(\p_n;s)$. \cref{fig:Acphic_3d} shows its first three modes extracted from Monte Carlo simulations, plotted as functions of $s$ alongside the $\cZ$ function at LPA and DE2. 
The lowest modes obtained from Monte Carlo simulations lie almost entirely within the DE2 result, up to error bars, and closely follow its amplitude and form, up to the statistical noise.  Note that the LPA result already gives the typical amplitude of $A(\p_n;s)$ around $s=0$, even though it cannot describe its magnetization dependence.
While strictly speaking $\cZ(s)$ corresponds to $A(\p_2;s)$ only, the fact that $A(\p_3;s)$ and $A(\p_4;s)$ are also well approximated by $\cZ(s)$ implies that $\cGamma(\p;s)$ has a quadratic dependence in $\p$ at small enough $\p$. (Recall that the dimensionful $\cGamma^{(2)}$ scales as $L^{-2+\eta}$, as expected.)

Together with our results for the rate function in three dimensions, this confirms that the DE2 expansion faithfully captures all significant features of the Monte Carlo data, and that the derivative expansion is a good approximation for the constraint effective action in three dimensions.

In two dimensions, \cref{fig:Acphic}, the Monte Carlo results show a rather surprising behavior in the second‐momentum component $A(\p_2,s)$, which departs markedly from the behavior of all higher‐momentum modes. The latter are more similar to the three-dimensional case, with a slow rise and fall with $s$. We find that the $\cZ$ function obtained from FRG at order DE2 is in relatively good agreement with the $A(\p_n;s)$ for $n>2$. We have checked that oscillations observed in $A(\p_2;s)$ are genuine, and not due to finite-size corrections. Indeed, the inset of \cref{fig:Acphic} shows that these oscillations obey scaling, as observed by changing the system size. It is still unclear what the cause is for these oscillations, but we hypothesize that they might be due to droplet effects induced by the fixed magnetization, which introduces an additional length in the system. Indeed, a similar, and much more pronounced, effect can be found in the one-dimensional Ising model in its scaling limit ($T\to0$ and $L\to\infty$), where a fixed magnetization imposes the existence of domain walls, and thus an additional length-scale associated to the domains that induces oscillations \cite{balog_constraint_2025}.

\section{Conclusion \label{sec:concl}}

In this paper, we have developed a robust framework to compute the constraint effective action at criticality using the FRG. Our approach extends beyond the LPA by incorporating a second-order derivative expansion, significantly enhancing the accuracy of the results. This is particularly evident in two-dimensional systems where the anomalous dimension plays a crucial role. Our results demonstrate excellent agreement with Monte Carlo simulations, validating the reliability and predictive power of the FRG method for computing universal scaling functions, including the rate function and momentum-dependent correlations at critical points.

We have also shown that the momentum dependence of the constraint vertices is rather non-trivial and differs from the usual behavior in the thermodynamic limit. Taking the example of the two-point constraint vertex, we have shown that it involves a new function $\cDelta_2(s)$ which vanishes in the thermodynamic limit. This is also the case for higher vertices, with additional functions that need to be studied in more detail.
Finally, it would be particularly insightful to apply the Blaizot--Mendez-Galain--Wschebor (BMW) approximation \cite{Blaizot2006,Benitez2012} within this framework to explore the full momentum dependence of the constraint correlation functions because it could clarify the subtle field and momentum behaviors observed, particularly in lower dimensions.

\acknowledgments
The authors thank B. Delamotte and N. Wschebor for useful discussions.
FR and AR thank the Institute of Physics of Zagreb for its hospitality. IB and AR acknowledge the support of the Croatian Science fund project HRZZ-IP-10-2022-9423. IB wishes to acknowledge the support of the INFaR and FrustKor
projects financed by the EU through the National Recovery and Resilience Plan (NRRP) 2021-
2026. I.B. acknowledges support from the project “Implementation of cutting-edge research and its application as part of the Scientific Center of Excellence for Quantum and Complex Systems, and Representations of Lie Algebras”, Grant No. PK.1.1.10.0004, co-financed by the European Union through the European Regional Development Fund - Competitiveness and Cohesion Programme 2021-2027. AR has benefited from the financial support of the Grant No. ANR-24-CE30-6695 FUSIoN, and acknowledges the support of the CDP C2EMPI, as well as the French State under the France-2030 programme, the University of Lille, the Initiative of Excellence of the University of Lille, the European Metropolis of Lille for their funding and support of the R-CDP-24-004-C2EMPI project.
This work was partially supported by an IEA CNRS project and by the “PHC COGITO” program (project number: 49149VE), funded by the French Ministry for Europe and Foreign Affairs, the French Ministry for Higher Education and Research, The Croatian Ministry of Science and Education, and the Deutsche Forschungsgemeinschaft (DFG, German Research Foundation) under Germany’s Excellence Strategy EXC2181/1-390900948 (the Heidelberg STRUCTURES Excellence Cluster). 

\appendix

\section{Momentum structure of the vertices}
\label{app:vertex}

As discussed in \cref{sec:finite_size}, at finite size, the two-point vertex in constant field is expected to take the form
\begin{equation}
	\cGamma_k^{(2)}(\p;s) = \cU''_k(s)+(1-\delta_{\p,\zero})\cDelta_{2,k}(s) + \cSigma_{2,k}(\p;s),
	\label{eq:2point}
\end{equation}
where $\cSigma_{2,k}(\p)$ includes the finite-momentum variation of the vertex, with $\cSigma_{2,k}(\zero)=0$, while $\cDelta_{2,k}(s)$ corresponds to a ``finite-momentum shift'' compared to the zero-momentum part $\cU''_k(s)$. As explained in \cref{sec:DE2}, a derivative expansion approximation of $\cSigma_{2,k}(\p)$ is then $\cSigma_{2,k}(\p)\simeq\cZ_k(s)\p^2+\ldots$

In the present Appendix, we discuss the momentum structure of the $n$-point functions at finite size, which differs from that in the thermodynamic limit. For this, it is convenient first to rewrite \cref{eq:2point} as\footnote{In this Appendix only, $\{\p_i\}_{i=1}^n$ correspond to $n$ arbitrary momenta indexed by the letter $i$, rather than to specific momenta $2\pi /L(i,0,\dots,0)$ along some direction, as in the other Sections.}
\begin{equation}
	\label{eq:2vertex_Gc_ansatz}
	\cGamma_k^{(2)}(\p_1,\p_2;s) =\delta_{\p_1+\p_2,\zero}\left( \cU''_k(s)+K_2(\p_1,\p_2)\cDelta_{2,k}(s) + \cSigma_{2,k}(\p_1,\p_2;s)\right).
\end{equation}
Here we have written the two-point vertex function as depending on two momenta, with the conservation of momentum included in the right-hand side (i.e. $\cGamma_k^{(2)}(\p;s)=\cGamma_k^{(2)}(\p,-\p;s)$), and we have introduced the symbol $K_i(\p_1,\ldots,\p_n)$, which is equal to $1$ if exactly $i$ of the $n$ momenta in its argument are not zero, and vanishes otherwise. For instance, $\delta_{\p_1+\p_2,\zero}K_2(\p_1,\p_2)=\delta_{\p_1+\p_2,\zero}(1-\delta_{\p_1,\zero})$.

With this notation, the three-point vertex function can thus be written, in analogy with \cref{eq:2vertex_Gc_ansatz}, as 
\begin{equation}
	\cGamma_k^{(3)}(\p_1,\p_2,\p_3;s) =\delta_{\p_1+\p_2+\p_3,\zero}\left( \cU^{(3)}_k(s)+K_3(\p_1,\p_2,\p_3)\cDelta_{3,k}(s)+K_2(\p_1,\p_2,\p_3)\partial_s\cDelta_{2,k}(s) + \cSigma_{3,k}(\p_1,\p_2,\p_3;s)\right).
	\label{eq:3point}
\end{equation}
Here $\cSigma_{3,k}(\p_1,\p_2,\p_3;s)$ is constrained when one momentum vanishes, e.g.  $\cSigma^{(3)}(\p_1,\p_2,\zero;s)=\partial_s \cSigma_k^{(2)}(\p_1,\p_2;s)$, which in addition to the term proportional to $K_2$ ensures that the exact relationship $\cGamma^{(3)}(\p_1,\p_2,\zero;s)=\partial_s \cGamma_k^{(2)}(\p_1,\p_2;s)$ is obeyed. (All vertices are invariant under exchange of momenta, and the corresponding relationships when a momentum is put to zero are assumed.)
Here again, $\cSigma_{3,k}(\p_1,\p_2,\p_3;s)$ is assumed to vanish when all its momenta are zero and to have a derivative expansion, with $\cDelta_{3,k}$ playing the role of a shift. Note that the two different shifts $\cDelta_{3,k}$ and $\partial_s\cDelta_{2,k}(s)$ never contribute at the same time, since their respective $K_n$ symbols are never non-zero at the same time. 

This can be generalized to an arbitrary $n$-point vertex as
\begin{equation}
	\cGamma_k^{(n)}(\{\p_i\}_{i=1}^n;s) =\delta_{\sum_{i=1}^n\p_i,\zero}\left( \cU^{(n)}_k(s)+\sum_{l=2}^nK_l\left(\{\p_i\}_{i=1}^n\right)\partial_s^{n-l}\cDelta_{l,k}(s)+\cSigma_{n,k}\left(\{\p_i\}_{i=1}^n;s\right)\right),
	\label{eq:npoint}
\end{equation}
with once again $\cSigma_{n,k}(\p_1,\ldots,\p_{n-1},\zero;s)=\partial_s\cSigma_{n-1,k}(\p_1,\ldots,\p_{n-1};s)$.
Note that the symbol $K_1$ never appears, since $\delta_{\sum_{i=1}^n\p_i,\zero}K_1\left(\{\p_i\}_{i=1}^n\right)$ always vanishes.

Importantly, in a derivative expansion, the various $\cSigma_{n,k}$ are not independent from each other, as happens in the thermodynamic limit. For instance, to order two in momenta, we have $\cSigma_{2,k}(\p_1,\p_2;s) = \cZ_k(s)(\p_1^2+\p_2^2+\p_1.\p_2)$, and consistency between vertices implies
\begin{equation}
	\begin{split}
		\cSigma_{3,k}(\p_1,\p_2,\p_3;s) &= \cZ_k'(s)(\p_1^2+\p_2^2+\p_3^2+\p_1.\p_2+\p_2.\p_3+\p_1.\p_3),\\
		\cSigma_{4,k}(\p_1,\p_2,\p_3,\p_4;s) &= \cZ_k''(s)(\p_1^2+\p_2^2+\p_3^2+\p_4^2+\p_1.\p_2+\p_2.\p_3+\p_3.\p_4+\p_1.\p_3+\p_1.\p_4+\p_2.\p_4).\\
	\end{split}
\end{equation}
However, no such relationships exist between the $\cDelta_{n,k}$, which are a priori independent functions.

In fact, it is possible to show that due to the constraint, the derivative expansion does not allow for closing the flow equations in terms of a finite number of functions ($\cU_k(s)$, $\cZ_k(s)$, $\cDelta_{2,k}(s)$, etc.). The simplest way to show this is to consider the flow at order DE2 of some arbitrary vertex function when all momenta are finite (otherwise, it is related directly to a lower-order vertex). It reads
\begin{equation}
	\partial_k \cGamma^{(n)}_{k}(\p_1,\ldots,\p_n;s) = \frac{1}{2L^d}\sum_{\q\neq\zero}\partial_k\cG_{k}(\q;s)\cGamma^{(n+2)}_{k}(\p_1,\ldots,\p_n,\q,-\q;s)+\ldots,
\end{equation}
where we only keep the simplest diagram to make our point.
Since all momenta on the right-hand side are finite, only $\cDelta_{n+2,k}(s)$ and $\cZ^{(n)}$ enter the vertex $\cGamma^{(n+2)}_{k}$. 

A similar argument shows that in every diagram, all vertices are evaluated at finite momenta, and a vertex $\Gamma^{(m)}_k$ will contribute to the flow through $\cDelta_{m,k}$ and $\cZ_k^{(m)}$. Thus, the flow of $\cDelta_{n,k}$ depends on $\cDelta_{n+1,k}$ and $\cDelta_{n+2,k}$, and the hierarchy does not close. In principle, one should compute the flow of all $\cDelta_{n,k}$, even within a finite order of the derivative expansion.

To close the flow equations, we assume that we can approximate $\cDelta_n(s)$ by $\partial_s^{n-2}\cDelta_2(s)$, which amounts to assuming that the finite-momentum shift does not depend on which momenta are finite or vanishing in \cref{eq:npoint}, i.e., that the finite momentum shift of $\cGamma^{(n)}$ is given for all momenta by $\partial_s^{n-2}\cDelta_2(s)$.

\section{Implementation of the FRG approach}
\label{app:FRG}

In this Appendix, we discuss some technical and numerical aspects of the FRG. 
The FRG approach involves the introduction of an infrared regulator function $R_k(\q)$, see \cref{eq:Wetterich}. In the present work, we use the Exponential regulator
\begin{equation}
\label{eq:exp_reg}
    R_k(\q)=\alpha Z_{k} k^2  e^{-\q^2/k^2}
\end{equation}
which has been used successfully to study DE to high orders~\cite{Balog2019}.
Here $Z_{k}$ is a scale of field renormalization needed to recover the critical physics (see below) and  $\alpha$ is a numerical prefactor. While the results should not, in principle, depend on $R_k(\q)$, any approximation induces a weak regulator dependence. We choose the value of $\alpha$ such that the results depend the least on the regulator, a procedure known as the 
principle  of minimal sensitivity (PMS) \cite{canet03,Kos2016}. The PMS is necessary to obtain convergence of the results with the order of the DE~\cite{Balog2019}. In practice, we choose the values of $\alpha$ found in the literature for the PMS of the critical exponent $\nu$. Another rationale for selecting the optimal $\alpha$, based on a minimal breaking of conformal symmetry at the RG fixed point leads to a very similar value of $\alpha$ \cite{balog_conformal_2020}.

In order to explore physics near criticality, it is necessary to introduce dimensionless variables through rescaling by appropriate powers of $k$. Indeed, $k$ explicitly breaks the scale invariance at criticality. We thus define the dimensionless quantities (noted with tildas)
\begin{gather}
\begin{aligned}
    \tilde \p &= \p/k, &
    \tilde s &= Z_k k^{2-d}s,
&
    \tilde\cU_k(\tilde{s})&=k^{-d} \cU_k(s), &
    \tilde\cDelta_{2,k}(\tilde{s})&=Z_k^{-1}k^{-2} \cDelta_{2,k}(s), &
    \tilde\cZ_k(\tilde{s})&=Z_k^{-1} \cZ_k(s). &
\end{aligned}
\end{gather}
The number $Z_k$ is necessary to take into account the anomalous dimension $\eta$ of the field. It is defined by the renormalization condition $\tilde\cZ_k(0)=1$ and at criticality $\partial_t Z_k \propto k^{-\eta}$. In dimensionless variables, the flow reaches a fixed point as $k$ goes to zero when the initial condition is tuned to criticality. In practice, we tune the initial condition close to criticality, and first run the flow down to some mesoscopic scale $k^*$, taken to be small with respect to the ultraviolet scales (e.g. inverse lattice spacing) and at which the potentials are very close to their fixed point values. We  introduce the RG time $t=\ln(k/k^*)$. Length scales are measured in units of ${k^*}^{-1}$.

Numerically, we integrate the flow equations using the explicit Euler method with a RG time step of $10^{-4}$. For the field dependence of the potential, we change to the variable $\tilde\rho=\tilde s^2/2$ and use an evenly spaced grid of two hundred points between $\rho=0$ and $\tilde{\rho}_{\text{max}}$. We used the so-called strict implementation of DE2, where all terms of order four or more in momenta are dropped from the integrals defining the flow equations (e.g., $|\p|^4$, $|\q|^4$, $\q^2\p^2$). However, we keep products of $\cDelta_{2,k}'$, which essentially play the role of a modification of the effective potential, in terms of the flow equation involving two $\cGamma^{(3)}_k$. Results with those parameters are stable up to at least five digits.

We now discuss implementation details specific to the problem of determining the rate function and working with a system of a finite size $L^d$. 
To investigate critical physics, we take $L\gg {k^*}^{-1}$. 
The finite size of the system implies that the momenta are discrete: in dimensionless variables $\tilde\q = 2\pi\n/kL$ with $\n \in \mathbb{Z}^d$. In the flow equations \eqref{eq:flowcU} and \eqref{eq:flow_2vertex_check}, $\partial_t R_k(\q)$ acts as an ultraviolet regulator and only modes with $|\q| \lesssim c k$ (with $c$ some constant) contribute to the sums, which include a number $O((kL)^{d})$ of terms. At the beginning of the flow, $k \gg L^{-1}$ makes directly computing the sums not tractable numerically. Rather, we remark that the difference between two consecutive modes $\Delta q = 2\pi/kL$ is small and approximate sums with integrals, e.g.
\begin{equation}
\label{eq:sum_by_int}
    \frac{1}{(kL)^{d}}\sum_{\tilde{\q} \neq 0} f(\tilde{\q}) = \int  \frac{d^d\tilde{\qc}}{(2\pi)^d} f(\tilde{\qc}) - \frac{1}{(kL)^{d}} f(0)+o\bigg(\frac{1}{(kL)^d}\bigg).
\end{equation}
Since the initial conditions $\cU_{k=k^*}(\rho)$, $\cZ_{k=k^*}(\rho)$ and $\cDelta_{2,k=k^*}(\rho)=0$ are the same as the potentials $U_{k=k^*}(\rho)$ and $Z_{k=k^*}(\rho)$ in the thermodynamic limit, at the beginning of the flow ($kL \gg 1$) the flow equations are identical, up to $O(1/(kL)^d)$ terms.

As $k$ approaches zero the approximation \labelcref{eq:sum_by_int} is eventually not justified anymore and we revert back to computing the sums directly. We note $k_\text{sum}$ the value of $k$ at which we make the switch; we typically pick $k_\text{sum}L\sim 40$ corresponding to an error in approximating sums with integrals smaller than $10^{-12}$. 

Last, in order to determine the value of the rate function at finite magnetization of order $L^{-(d-2+\eta)/2}$, we switch back to a dimensionful field grid at the end of the flow. Indeed, the dimensionless grid $0 \leq \tilde \rho \leq \tilde{\rho}_\text{max} = k^{d-2+\eta} \tilde{\rho}_\text{max}$ shrinks around $0$ as $k$ is lowered. The change is made when $k$ is such that the largest dimensionless $\tilde s_\text{max}=\sqrt{2 \tilde \rho_\text{max}}$ corresponds to the desired dimensionful $s$, typically two times the position of the minimum.

\section{Dependence on the regulator parameter \texorpdfstring{$\alpha$}{α} \label{app:alpha}}

We discuss here the dependence of our DE2 results  as a function of the regulator parameter $\alpha$, see \cref{eq:exp_reg}. The dependence of the rate function on $\alpha$ within the LPA was discussed in Ref.~\cite{Balog2022}.

\begin{figure}[ht!]
    \centering
    \includegraphics[width=0.45\linewidth]{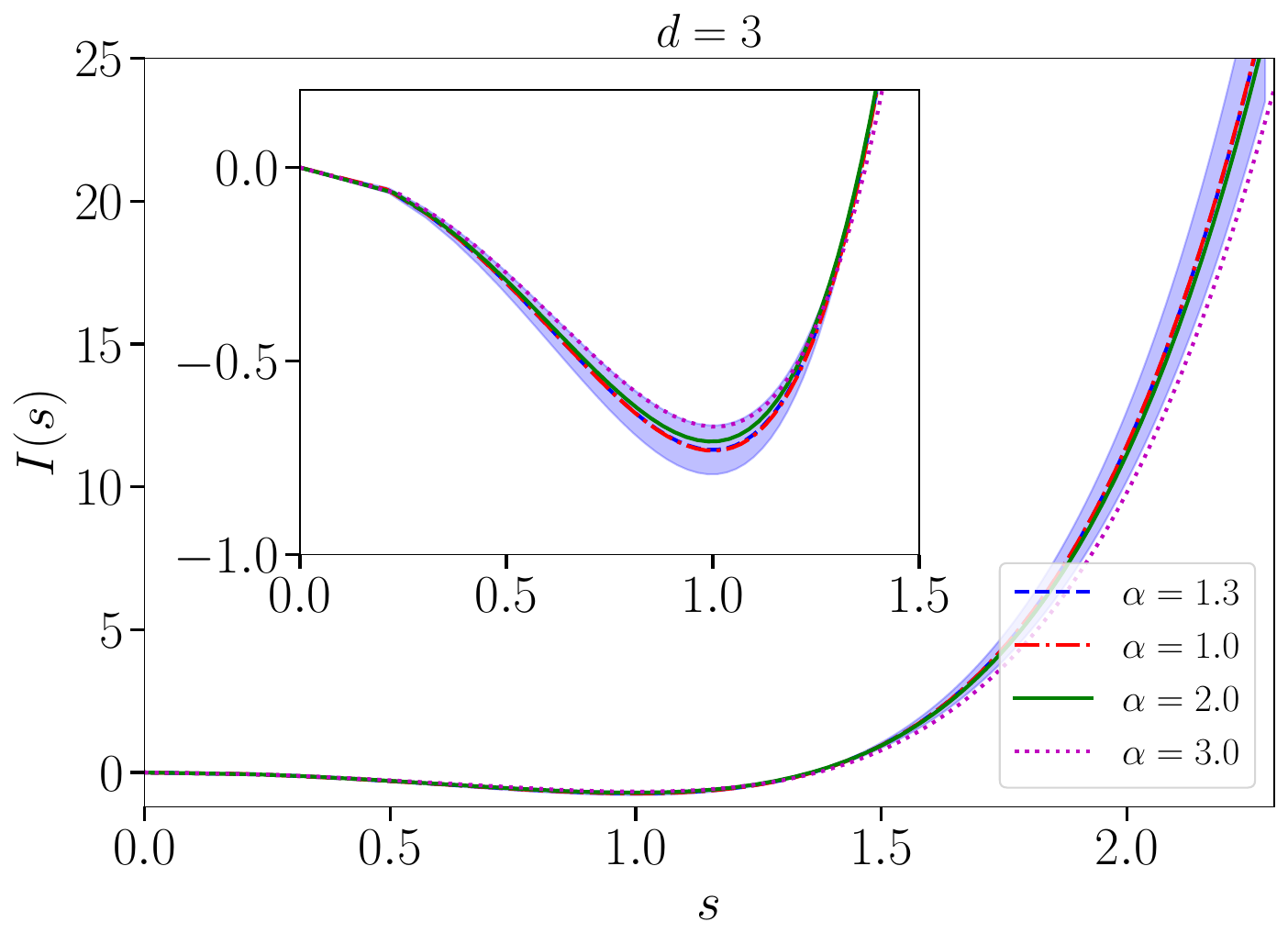}
    \includegraphics[width=0.45\linewidth]{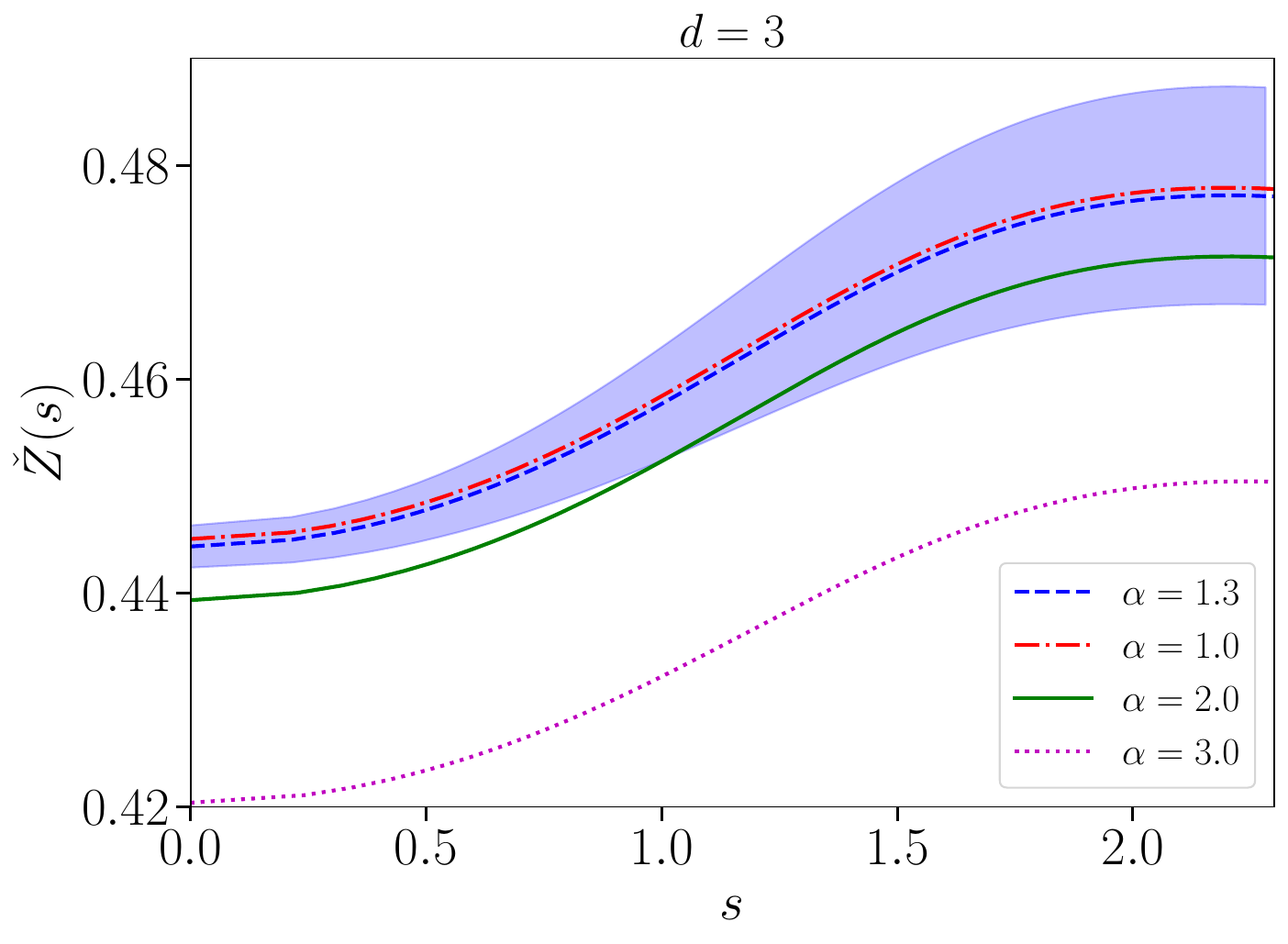}
    \caption{Rate function $I(s)$ (left panel) and function $\cZ(s)$ (right panel) as a function of $s$ in $d=3$ for various values of the regulator parameter $\alpha$. The optimal value used in the main text is $\alpha=1.3$. The shaded area represents the error bars associated to the truncation (see main text). The inset in the left plot is a zoom around the minimum of the rate function. The apparent change of slope at small $s$ is due to the discretization of $s$ on a grid.}
    \label{fig:alpha_3D}
\end{figure}

\Cref{fig:alpha_3D} shows $I(s)$ and $\cZ(s)$ as a function of $s$ in $d=3$ for various values of $\alpha$. The optimal value in this case is $\alpha=1.3$. For the rate function, we observe that the dependence on $\alpha$ is weak, as the results remain almost entirely within the error bars that come from the comparison with the LPA (see the main text) even for $\alpha$ as large as $3$. The dependence on $\alpha$ of $\cZ(s)$ is stronger, but we note first that the change is of the order of a few percent and second that it concerns mostly the amplitude, the overall shape being very similar even for $\alpha=3$.

\begin{figure}[ht!]
    \centering
    \includegraphics[width=0.5\linewidth]{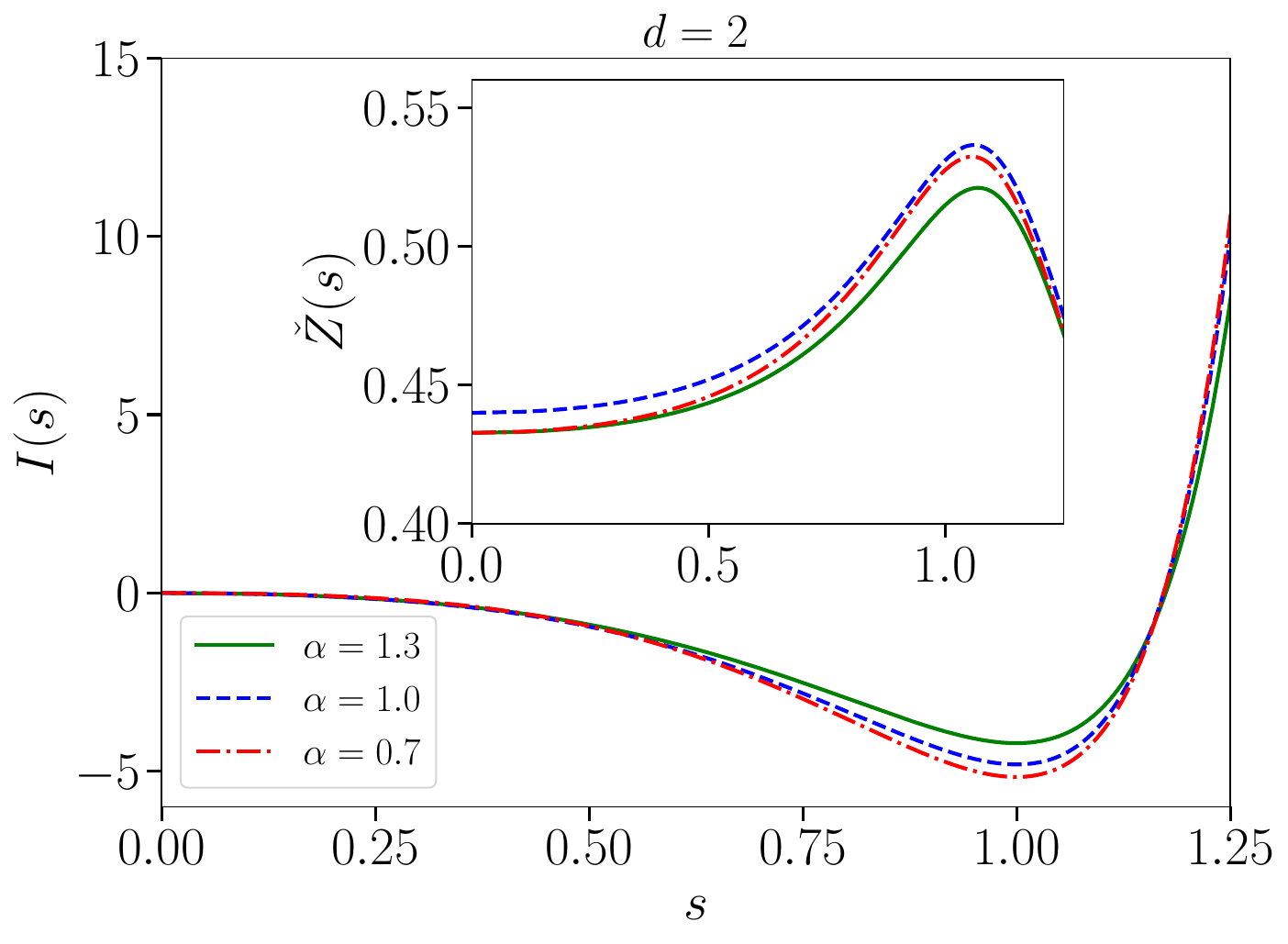}
    \caption{Rate function $I(s)$ (main panel) and function $\cZ(s)$ (inset) as a function of $s$ in $d=2$ for various values of the regulator parameter $\alpha$. The optimal value used in the main text is $\alpha=1.0$.}
    \label{fig:alpha_2D}
\end{figure}

In two dimensions, see \cref{fig:alpha_2D}, the dependence on $\alpha$ is stronger, as expected. The value of the minimum is rather sensitive to $\alpha$, while $\cZ(s)$ is much less so.

\end{document}